\documentclass[twocolumn,english,aps,superscriptaddress, pra,groupedaddress]{revtex4-1}

\usepackage[latin9]{inputenc}
\usepackage{amsmath}
\usepackage{amssymb}
\usepackage{graphicx}
\usepackage{babel}
\usepackage{mathrsfs}
\usepackage{amsfonts}
\usepackage{epstopdf}
\usepackage{caption}
\usepackage{multirow}

\begin{document}

\title{Quantum Many-Body Conformal Dynamics: \\ Symmetries, Geometry, Conformal Tower States, and Entropy Production }

\author{Jeff Maki and Fei Zhou}

\affiliation{ Department of Physics and Astronomy, University of British Columbia, Vancouver V6T 1Z1, Canada}

\date{\today}

\begin{abstract} 
In this article we study the quench dynamics of Galilean and scale invariant many-body systems which can be prepared using interacting atomic gases. The far-away from equilibrium dynamics are investigated by employing $m$-body density matrices, which are most conveniently defined in terms of a special basis - the conformal tower states. We explicitly illustrate that, although during the initial stage of the dynamics all symmetries can be broken and absent in the unitary evolution because of the initialization of the state, there is always an emergent conformal symmetry in the long time limit. The emergence of this dynamic conformal symmetry is robust, and always occurs - even when scale and other symmetries (such as rotational symmetry) are still fully broken in the many-body states; it uniquely defines the characteristics of the asymptotic dynamics near a scale invariant strong coupling fixed point. As an immediate application of the asymptotic dynamics of the microscopic density matrices, we have focused on the effects of this emergent conformal symmetry on two observables: the moment of inertia tensor, $I_{ij}(t)$, $i,j=x,y,z$, and the entropy density field, $S({\bf r}, t)$, in the hydrodynamic flow of strongly interacting particles. We show that the long time behaviour of these observables is completely set by conformal symmetry, while the leading long time corrections depend on interference effects between different conformal tower states. The emergent conformal symmetry naturally leads to entropy conservation, and {\em conformal cooling}, an energy conserving cooling of a strongly interacting gas during free expansion. When the interaction Hamiltonian breaks the scale symmetry, we further demonstrate that there is a direct cause-effect relation between conformal symmetry breaking in the long time limit, and a non-vanishing entropy production. This suggests that the entropy production rate is a natural {\em parameter} for categorizing the breaking of conformal symmetry.
\end{abstract}

\maketitle

\section{Introduction}

Far-from equilibrium quantum phenomena in the limit of strong interactions have attracted enormous attention in recent years. To fully investigate quantum dynamics, it is necessary to have precise control over a quantum system in order to prepare a wide range of interactions and parameters for initialization of states. Thankfully, such a platform does exist; atomic gas systems. A variety of dynamical phenomena can be studied thanks to recent technological advancements that have led to the ability to masterfully control the atomic confining potentials, and mutual inter-particle interactions \cite{Bloch08,Chin10}.

A particularly interesting subclass of dynamical experiments is to perform a quantum quench to or at strong interactions. This has allowed for the study of phenomena like collective modes, expansion dynamics, non-linear dynamics, hydrodynamics, many-body instabilities in strongly interacting limits, and more \cite{O'Hara02, Kinast04, Bartenstein04, Spivak04, Kinast06, Bulgac09, Cao11, Bulgac12,Grimm13a, Grimm13b,Elliott14, Zwierlein16, Rem13, Makotyn14, Sykes14, Yin13, Jiang14, Jiang16, Eismann16, Rancon14,Chevy16, Eigen17}. Such situations naturally arise in atomic gases thanks to the control present in atomic gas systems. As we will discuss later on, one can show that the effective Hamiltonian for these strongly interacting systems may have additional symmetries consistent with Galilean invariance, such as scale and conformal symmetry, if the interaction Hamiltonian is tuned to a strong coupling fixed point which represents resonant unitary gases.

Scale and conformal symmetry, as we discuss below, are defined as the invariance of the equations of motion for many-body unitary evolution under the following re-parametrizations of the spatial and temporal coordinates:

\begin{align}
{\bf r}_i' &= {\bf r}_ie^{-\lambda} & t' &= t e^{- 2 \lambda},
\label{eq:scale_inv}
\end{align}

\noindent for scale symmetry, and:

\begin{align}
{\bf r}_i' &= \frac{{\bf r}_i}{1- \lambda t} & t' &= \frac{t}{1- \lambda t},
\label{eq:conf_inv}
\end{align}

\noindent for conformal symmetry \cite{Hagen72, Niederer72, Henkel94, Wingate06,Son07}. In Eqs.~(\ref{eq:scale_inv}) and (\ref{eq:conf_inv}), ${\bf r}_i$ is the position of the $i = 1,2,...,N$ particle in the atomic gas, and $\lambda$ parametrizes the extent of the transformation. These two symmetries can drastically reduce the complexity of the dynamics, and potentially allow one to understand the dynamics of strongly interacting systems, which are often theoretically intractable due to the lack of a small parameter.

A thorough study of the role of scale and conformal symmetries on the quench dynamics of atomic gases is indispensable as it will also shed light on the general properties of non-equilibrium phenomena near quantum critical points that exhibit the same symmetries. For example, these symmetries (Galilean, scale, and conformal invariance) can also occur in a much broader class of solid state systems near Lifshitz transitions around the bottom of a band with aysmptotic quadratic dispersions, where the Fermi surface topology undergoes a drastic change due to varying external parameters \cite{Lifshitz60}. Equally important, the results of such a study  can serve as a starting point for future studies of more generic strongly interacting physical systems that break these symmetries; i.e. one can study the dynamics of symmetry broken systems by comparing to their symmetric counterpart.

Below we are mainly interested in exploring possible experimental signatures of scale and conformal symmetries. For this special class of scale and conformal invariant Hamiltonians, a number of peculiar aspects have been investigated in quantum dynamics. The earliest attempt to connect conformal symmetry to the dynamics of cold gases was presented in Ref.~\cite{Rosch97}. In this work it was argued that the breathing modes of a two dimensional Bose gas in an isotropic harmonic trap would be at exactly twice the trap frequency. Now it is known this statement is only approximately true, as the two-dimensional Bose gas is not scale invariant due to the quantum anomaly \cite{Olshanii10, Hofmann12, Gao12}. Nevertheless, their result applies to any scale invariant system placed in an isotropic harmonic trap.

Later, it was experimentally demonstrated that the damping of collective modes for a strongly interacting three-dimensional Fermi gas will have a minimum right at resonance where one anticipates interactions are scale symmetric \cite{Bartenstein04}. This is a pleasant surprise because the scattering cross section actually reaches a maximum or infinite value at resonance which classically would have led to a maximum in the damping rate. Moreover, this situation occurs in the BEC-BCS crossover  regime where there are no physical phase transitions when varying magnetic fields. 

Another important consequence of scale and conformal symmetry is related to the shear and bulk viscosity of an atomic gas. For hydrodynamics that are characterized by a set of general-coordinate and conformal invariant equations, it was theoretically demonstrated that the bulk viscosities will vanish identically in the pioneering work of D. T. Son \cite{Son07}. Other approaches either based on the conventional perturbative diagrammatic calculations, or the general sum rules also suggest a consistent picture \cite{Enss11,Taylor12}. 

Certain aspects of the hydrodynamics were investigated experimentally for three dimensional unitary Fermi gases by Thomas' group \cite{O'Hara02,Cao11,Elliott14}. In their experiments, they were able to measure both the shear and bulk viscosities by examining the expansion dynamics in the presence of resonant interactions. The expansion dynamics were then modelled using a scaling ansatz consistent with the hydrodynamic equations of motion \cite{O'Hara02, Cao11, Elliott14, Stringari02}. The results of this experiment are consistent with the predictions of Ref.~\cite{Son07}.

Although the variational solutions to the hydrodynamic approach are in good agreement with the experimental data, it does rely on important inputs of phenomenological parameters that can only be obtained via other microscopic considerations, or sometimes by a separate analysis of the implications of general scale and conformal symmetries. It further relies on the existence and knowledge of {\em thermodynamic-like} equations of state in far-away from equilibrium quantum phenomena. Even with various inputs from other considerations being available, a full simulation of flow fields in most generic situations is very challenging, and severely restricted by the current available computational power. Practically, one usually has to introduce a very specific empirical ansatz to model the hydrodynamic flow and compare with experimental data.

Evidently, there needs  to be a transparent first-principle-based microscopic view of why strongly interacting many body systems behave in such highly surprising ways. This can be achieved by a density-matrix based theory which effectively closes the extensive gap between hydrodynamic phenomenologies and microscopic unitary evolution of quantum many-body states. This is one of the main objectives to achieve in this article.

More importantly, we address the question: in general scale invariant critical phenomena, broadly speaking, what is the dynamical consequence of the additional conformal symmetry and its breaking? The possibility of performing controllable quench experiments with these dynamic symmetries further raises a few more unique and fundamental questions.

1) In dynamics, generically all the symmetries of the Hamiltonian can be broken by initial conditions. A reasonable question to then ask is: {\em which symmetries, if any, can prevail in the asymptotic long time dynamics?}  Are both scale and conformal symmetries re-emergent in the long time dynamics, or are these two symmetries mutually exclusive and only one of them will emerge? 

2) Secondly: {\em if such an emergent symmetry does exist, what are the experimental consequences of this symmetry on the expansion dynamics of scale invariant systems?} 
Are the implications consistent with the hydrodynamical flow studied experimentally before, and do the emergent symmetries suggest new features that  haven't been fully  understood or observed previously?

3) Third: {\em whether there is an explicit relation between the prevailing/absence of space-time scale and conformal symmetries and entropy conservation/ production}. This issue lies at the heart of  hydrodynamics and is connected to the bulk viscosity. Since these two symmetries imply a vanishing bulk viscosity, can the entropy production rate be a natural symmetry breaking {\it parameter} which characterizes the breaking of the emergent conformal symmetry?

In this article, we address these questions by showing that for very generic initial conditions, where all the symmetries can be broken during the early stages of the unitary evolution, the asymptotic dynamics of scale invariant Hamiltonians, $H_s$, is always governed by an emergent conformal symmetry - while scale and other symmetries are still fully broken in the many-body state. The unique role of conformal symmetry in the asymptotic dynamics can be seen by considering the transformation in Eq.~(\ref{eq:conf_inv}).  For scale invariant interactions (defined in Sec.~\ref{sec:scale_conf}), the unitary evolution of the fermion field operator, $\psi_{\sigma}({\bf r},t)$, is governed by 

\begin{equation}
\partial_t \psi_{\sigma}({\bf r},t) = i \left[ H, \psi_{\sigma}({\bf r},t) \right].
\label{eq:Heisenberg_eqn}
\end{equation}

\noindent Eq.~(\ref{eq:Heisenberg_eqn}) remains invariant in the transformed space-time defined by Eq.~(\ref{eq:conf_inv}). That is, the properly transformed field operator, $\psi_{\sigma}({\bf r'},t')$, obeys the identical dynamic equation, with the same scale invariant Hamiltonian, $H_s$, re-expressed in terms of the field-operator: $\psi_\sigma({\bf r'}, t')$.

\begin{figure*}
\includegraphics[scale=0.35]{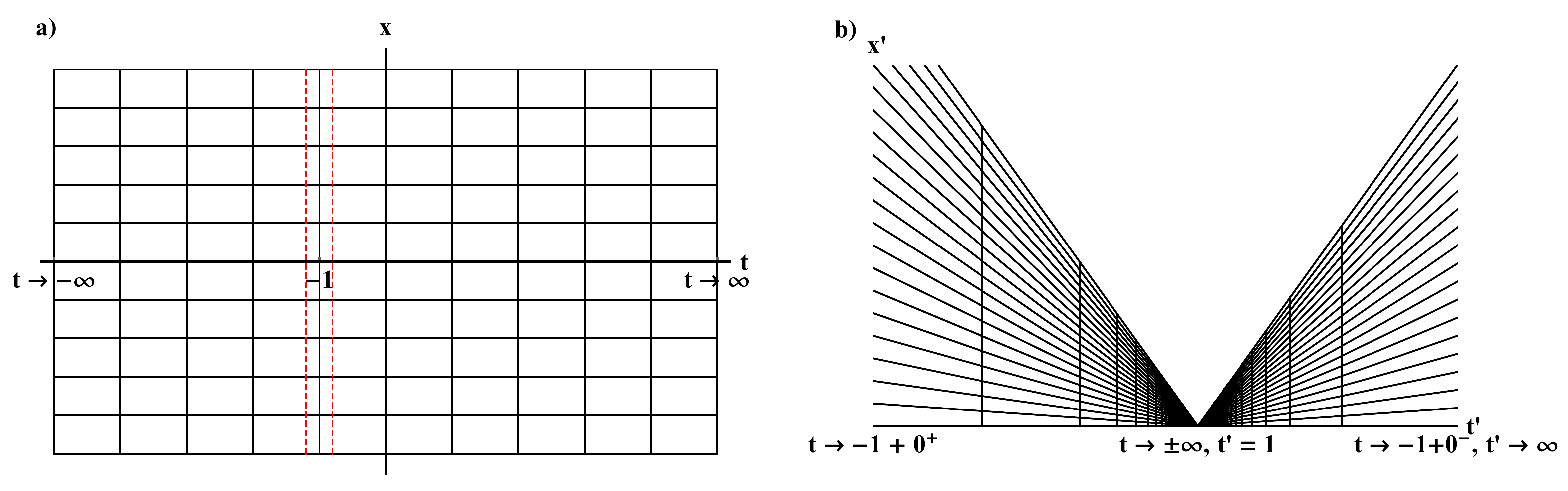}
\caption{The action of conformal transformation on a space-time grid. 
a) Horizontal (vertical) lines in the grid represent equal-space (equal-time) lines with constant intervals in space (time), before the transformation. b) The space-time grid after the conformal transformation with $\lambda$ in Eq.~(\ref{eq:conf_inv}) set to be $-1$. This geometry, defined by Eq.~(\ref{eq:conf_inv}) can be obtained by cutting the space-time grid along the $t = -1$ (see the red dashed lines in a)) line to create two separate time domains $[-\infty, -1-0^+]$ and $[-1+0^+, \infty]$, and then swapping the two domains - transporting $t=-1+{0^+}$ to $t'=-\infty$ and $t=-1+{0^-}$ to $t'=+\infty$. This results in $t=+\infty$ (now $t'=1-0^+$) being to the left of $t=-\infty$ line (now as $t'=1+0^+$). As one can see, all the equal-x lines in a) are converted into straight lines emitted from the point $(t' =1, x' = 0)$. Consequently, all the long time dynamics $t \rightarrow \pm \infty$ are compactified near $t' \approx 1$ in the transformed space-time geometry.}
\label{fig:geometry}
\end{figure*}

In Fig.~(\ref{fig:geometry}), we illustrate the relation between the original space-time geometry and the conformally transformed space-time structure defined in Eq.~(\ref{eq:conf_inv}). The important feature is that the whole spatial space at $t=\infty$ is completely compactified into the single point $(x'=0,t'=1)$.  This effectively converts all the long time dynamics of the quantum system at $t \rightarrow \infty$, into the equivalent dynamics in the vicinity of $t' =1$, in the transformed geometry. This drastic compactification offers a simple qualitative picture of the long time asymptotic dynamics. Namely, since under the transformation the scale invariant Hamiltonian, $H_s$, remains invariant, it {\em does not}  depend on $t'$ explicitly  after the transformation. The unitary evolution in the vicinity of $t'=1$ in the transformed coordinates effectively has little dependence on $(1-t')$, the distance from $t'=1$, as $U(t')=\exp(i H_s t') \approx U(t'=1) +O(1-t')$. Since the unitary evolution {\em freezes out} near $t'=1$ or $t \rightarrow \infty$, the dynamics of local observables are simply related to reparameterization of the spatial coordinates suggested by the transformation given in Eq.~(\ref{eq:conf_inv}). It then follows that properties such as the total thermodynamic entropy will saturate, or equivalently, the entropy production rate will rapidly approach zero.

We shall remark that this line of argument depends crucially on the underlying scale symmetry of the Hamiltonian, and thus is only valid if and only if we are dealing with a scale invariant fixed point Hamiltonian. If the Hamiltonian is not scale invariant, and the interactions deviate from the fixed point value, the Hamiltonian in the transformed coordinates, $H'(t')$, will have explicit dependence on $t'$. There are two situations that can occur. If the unitary evolution $U(t')$, or the action of Hamiltonian $H'(t')$, is still analytical near $t' =1$, then $U(t') \approx U(t'=1)+O(1-t')$, and the long time dynamics will still be constrained by an emergent conformal symmetry. However, if instead the action of the time dependent Hamiltonian, $H'(t')$, is singular near $t'=1$ and $U(t')$ is not analytic at $t'=1$,  then the effect of the deviation from the scale invariant fixed point is relevant in the long time dynamics, as a small change in $t'$ near $t'=1$ is expected to result in a non-perturbative change in the unitary evolution.  Below we will discuss consequence of the non-perturbative singular effects in the context of elliptic flow and entropy production.  One can find related more discussions on classifications of irrelevant and relevant operators via the beta-function method in Ref.~\cite{Maki18}.

In this article we prove that there is an emergent conformal symmetry, and show that our geometrical description is valid by means of exact symmetry arguments and the $m$-body density matrix. This approach is microscopic and will clearly elucidate the features of conformal and scale symmetry on the long time dynamics of quantum systems with scale invariant Hamiltonians. The remainder of this article is organized as follows. 

In section \ref{sec:scale_conf}, we review the general concepts of scale and conformal transformations, and define the respective symmetries associated with these transformations. In addition, we introduce the concept of conformal tower states, which are a set of many-body states whose temporal evolution is equivalent to a time dependent rescaling, and whose existence is guaranteed by the scale and conformal symmetries. In section \ref{sec:den_matrix}, we construct the density matrices for quench dynamics utilizing  these conformal tower states. In Sec. \ref{sec:re_emergence}, we show that although the initial quantum state can break all of the symmetries, rotational, scale, and conformal, we find that at long times, the density matrix will be asymptotically conformal invariant, or equivalently, will be an eigenfunction of the generator of conformal transformations with zero eigenvalue. In Sec \ref{sec:breaksi}, we discuss how the dynamics of the density matrices are modified by explicitly breaking the scale invariance of the Hamiltonian. In this case, we focus on Hamiltonians that are singular near $t' \rightarrow 1$, as the dynamics of such a system are non-trivial compared to their scale invariant counterpart. 

We follow these formal discussions with a practical investigation of the hydrodynamic flow of the three-dimensional unitary Fermi gas, Sec.~\ref{sec:hydro_flow}. We show that the elliptic and compressional hydrodynamic flows depend on the inter- and intra-conformal tower interference, respectively. Our microscopic approach is then compared to the general hydrodynamics phenomenologies discussed previously \cite{Cao11,Elliott14, Grimm13b, Stringari02}.

We finish our discussions in Sec.~\ref{sec:ent} by examining the time evolution of the thermodynamic entropy. In particular, we show there is a one-to-one correspondence between the entropy production and the explicit breaking of the emergent conformal symmetry due to a deviation of the interaction from  its scale invariant value. Our studies hence suggest that the entropy production rate can be used to parametrize the breaking of conformal symmetry. We then present our final conclusions in Sec.~\ref{sec:conclusions}.

\section{Scale and Conformal Symmetry}
\label{sec:scale_conf}

We begin our discussions by reviewing the definitions of scale and conformal transformations. In quantum mechanics, a continuous transformation is performed by means of a unitary operator:

\begin{equation}
U_G(\lambda) = e^{-i G \lambda},
\end{equation}

\noindent where $\lambda$ is a generalized angle that parametrizes the magnitude of the transformation, and $G$ is a Hermitian operator known as the generator of the transformation. For scale and conformal transformations, the generators are respectively given by:

\begin{align}
D &=\int d{\bf r} \  \psi_{\sigma}^\dagger({\bf r}) [-i\frac{d}{2} +{\bf r}\cdot {\bf p}]\psi_{\sigma}({\bf r}) \nonumber \\ 
&C=\int d{\bf r} \ \psi_{\sigma}^\dagger ({\bf r}) \frac{r^2}{2}\psi_{\sigma}({\bf r}),
\end{align}

\noindent where $\psi_{\sigma}({\bf r})$ is the field operator for a fermion with spin, $\sigma$, which satisfies the anti-commutation relation:

\begin{equation}
\left \lbrace \psi^{\dagger}_{\sigma}({\bf r}), \psi_{\sigma'}({\bf r'}) \right \rbrace = \delta_{\sigma, \sigma'} \delta({\bf r - r'}).
\end{equation}

Under these transformations, the field operator transforms as:

\begin{align}
U_D(\lambda) \psi_{\sigma}({\bf r}) U_D^{\dagger}(\lambda) &= e^{- \lambda d/2} \psi_{\sigma}( e^{-\lambda} {\bf r}), \nonumber \\
U_C(\lambda) \psi_{\sigma}({\bf r}) U_C^{\dagger}(\lambda) &= e^{-i {\bf r}^2/2} \psi_{\sigma}({\bf r}), \nonumber \\
\end{align}

\noindent where $d$ is the dimension of the system.

In order to see how these transformations affect the dynamics, it is necessary to consider how the time evolved field operators transform under scale and conformal transformations. To do this, consider a system of fermions with short ranged s-wave interactions in $d$ spatial dimensions. This system can be accurately modelled by the following Galilean invariant effective theory:

\begin{align}
H[g(\Lambda), \Lambda] &= \sum_{\sigma} \int_{\Lambda} d{\bf r} \ \psi^\dagger _{\sigma} ({\bf r})\left(\frac{p^2}{2} \right)  \psi_\sigma(\bf r)  \nonumber \\
&+\frac{1}{2} g(\Lambda)  \sum_{\sigma}\int_\Lambda d{\bf r} \ \psi^\dagger_\sigma ({\bf r}) \psi^\dagger_{-\sigma} (\bf r)
\psi_{-\sigma}({\bf r}) \psi_{\sigma}(\bf r) ,
\label{eq:Hamiltonian}
\end{align}

\noindent where the Hamiltonian is specified by $\Lambda$, the ultra violet cut-off, and $g(\Lambda)$, the corresponding interaction constant. Here we note that all integrations are restricted such that $r > \Lambda^{-1}$. For the remainder of the discussion, the spin indices will not play an important role, and will be suppressed.

Before one can discuss how scale and conformal transformations affect the dynamics of the field operator, it is important to note that this Hamiltonian is renormalizable.  One can formulate the theory at different scales, $\Lambda$, without changing the physical properties of the system, if we rescale $g(\Lambda)$ accordingly. The change in the Hamiltonian, or equivalently $g(\Lambda)$, is characterized by the renormalization equation (RG) flow of the equivalent Hamiltonians, $H[g(\Lambda), \Lambda]$, defined at different scales, $\Lambda$ \cite{Wilson83, Sachdev}. 

The flow of the coupling constant is intimately connected to how the Hamiltonian changes under scale transformations. To describe the flow of the coupling constant, it is necessary to examine the dimensionless coupling:

\begin{equation}
\tilde{g}(\Lambda) =C_d g(\Lambda)\Lambda^{d-2},
\end{equation}

\noindent where $C_d$ is a constant that depends on the dimension, $d$. The change of the dimensionless coupling constant as a function of $\Lambda$ is given by the beta-function:

\begin{equation}
\frac{d \tilde{g}(\Lambda)}{d \ln \Lambda}=\beta(\tilde{g}(\Lambda)) = (d-2) \tilde{g}(\Lambda)+ \tilde{g}^2(\Lambda),
\label{eq:beta_function}
\end{equation}

\noindent where the second equality is the result for our effective model, Eq.~(\ref{eq:Hamiltonian}) \cite{Sachdev, Maki18}. If $\beta(\tilde{g}(\Lambda))$ vanishes, the coupling constant does not change under a rescaling of the ultra violet cut-off. At this so called fixed point, the system is scale invariant, as the Hamiltonian is invariant under a rescaling of $\Lambda \rightarrow e^{-
\lambda}\Lambda$. We denote the scale invariant value of the coupling constant as $\tilde{g}(\Lambda) = \tilde{g}^*$, and the scale invariant Hamiltonian as: $H[\tilde{g}^*,\Lambda] = H_s$.

In the context of cold gases, there are two scale invariant fixed points. These correspond to the non-interacting, and resonantly interacting fixed points. In terms of the dimensionless coupling constant, these two points respectively correspond to:

\begin{align}
\tilde{g}^* &= 0 & \tilde{g}^* &= 2-d.  
\end{align}

\noindent Both of these fixed points can be achieved thanks to the presence of a Feshbach resonance. 

\begin{figure*}
\includegraphics[scale=0.3]{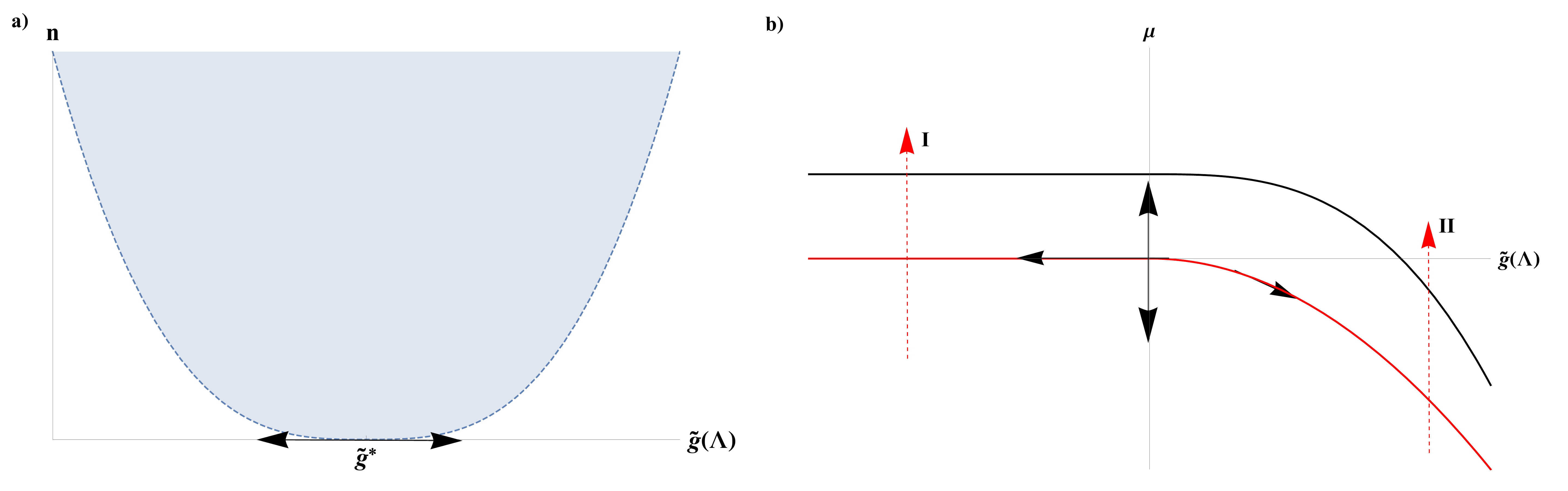}
\caption{Thermodynamics near a strong coupling fixed point. $n$ is the density, and $\tilde{g}(\Lambda)$ is the dimensionless interaction constant. a) The shaded region in the $n-\tilde{g}(\Lambda)$ plane represents the space where the thermodynamics is governed by the scale invariant point, $\tilde{g}^*$. In this region, the effect of breaking scale invariance is perturbative, and the thermodynamics are dictated by the scale invariance at $\tilde{g}^*$.  The non-shaded area represents a weakly interacting  system. For cold atoms this corresponds to a dilute gas of either atoms or molecules. The arrows along the $\tilde{g}(\Lambda)$ axis indicate the renormalization group flow out of the unstable fixed point $\tilde{g}^*$. For three dimensional cold gases, this corresponds to the unitary scale invariant fixed point, $\tilde{g}^*=-1$. b) A schematic of the RG flow of the chemical potential, $\mu$, and $\tilde{g}(\Lambda)$ around the point, $(\mu = 0,\tilde{g}(\Lambda) = \tilde{g}^*)$. The $n=0$ line in a) is mapped onto the red line in the $\mu-\tilde{g}$ plane. This line is the critical line, separating the vacuum from finite density quantum gas phases. The solid black line represents the smooth change of chemical potential of a quantum gas with a fixed density, as the interaction is increased from weak (left hand side) to strong (right hand sign) across the vertical axis of $\tilde{g}=\tilde{g}^*$. The transition along trajectory I belongs to free-fermion universality class, while trajectory II belongs to the free boson class. The point $(\mu=0, \tilde{g}^*)$ separates these two classes (and appears to be multiple critical in this plane). This point is the strong coupling fixed point which we focus on. It further exhibits $SO(2,1)$ conformal symmetry and dictates non-equilibrium dynamics in the near vicinity as discussed below.  }
\label{fig:td_coupling}
\end{figure*}

The stability of these fixed points to perturbations depends on the derivative of the beta function, $\beta'(\tilde{g}^*)$. If $\beta'(\tilde{g}^*)<0$, the perturbation is relevant, and it drives the system away from scale invariance as one lowers the UV cut-off. In the opposite limit, $\beta'(\tilde{g}^*) >0$, the perturbation is irrelevant, and the thermodynamics are governed by the scale invariant fixed point. In Fig.~(\ref{fig:td_coupling}), we illustrate these features near the resonant fixed point for the three dimensional Fermi gas, $\tilde{g}^* = -1$. The shaded region represents the area where the physics is governed by the scale invariant point, $\tilde{g}^*$, while the non-shaded region represents the area where the physics will deviate, and eventually flow into the other fixed point.

One important consequence of fine tuning the Hamiltonian to a scale invariant fixed point is that a hidden symmetry emerges. This hidden symmetry can be seen by noting that the commutators between the scale invariant Hamiltonian, $H_s$, and the generators of scale, $D$, and conformal, $C$, transformations form a closed algebra:

\begin{align}
[ H_s, C] &= -i D & [D, H_s] &= 2i H_s & [D, C] &= -2 i C.
\label{eq:SO21_algebra}
\end{align} 

\noindent These commutators are known as the conformal algebra, and form a representation of the group SO(2,1) \cite{Hagen72,Nishida07}.

Using the so(2,1) algebra, Eq.~(\ref{eq:SO21_algebra}), it is possible to understand how the time dependent field operator:

\begin{equation}
\psi_s({\bf r},t) = e^{i H_s t} \psi({\bf r}) e^{-i H_s t},
\end{equation}

\noindent transforms under scale and conformal transformations. The results for scale and conformal transformations, respectively, are given by:

\begin{align}
U_{D}(\lambda) \psi_s({\bf r},t) U^{\dagger}_{D}(\lambda) &=e^{-\lambda \frac{d}{2}} \psi_s({\bf r}'={\bf r}e^{-\lambda}, t'=t e^{-2\lambda}), 
\nonumber \\ 
U_{C} (\lambda)  \psi_s({\bf r}, t) U^{\dagger}_{C}(\lambda) &=\frac{1}{(1-\lambda t)^{d/2}}\exp(-i\frac{r^2}{2}\frac{\lambda}{1-\lambda t})
\nonumber  \\
& \psi_s({\bf r}'=\frac{\bf r}{1-\lambda t}, t'=\frac{t}{1-\lambda t}). 
\label{eq:time_dependent_transformation}
\end{align}

\noindent As one can see, the transformed field operators, Eq.~(\ref{eq:time_dependent_transformation}), are consistent with Eqs.~(\ref{eq:scale_inv}) and (\ref{eq:conf_inv}).

Similarly, it is possible to show that the equation of motion:

\begin{equation}
\partial_t  \psi_s({\bf r},t) = i \left[H_s, \psi_s({\bf r},t)\right],
\label{eq:eom}
\end{equation}

\noindent is left invariant under scale and conformal transformations, if one uses the coordinates, $({\bf r'},t')$, defined in Eqs.~(\ref{eq:scale_inv}), and (\ref{eq:conf_inv}). To see this, consider:

\begin{align}
\partial_t U_C(\lambda) \psi_s({\bf r},t) U^{\dagger}_C(\lambda) &=  i U_C(\lambda) \left[ H_s, \psi_s({\bf r},t)\right]  U_C^{\dagger}(\lambda) \nonumber \\
\partial_t U_D(\lambda) \psi_s({\bf r},t) U^{\dagger}_D(\lambda) &=  i U_D(\lambda) \left[ H_s, \psi_s({\bf r},t)\right]  U_D^{\dagger}(\lambda).
\label{eq:transformed_eom}
\end{align}

\noindent With the aid of Eqs.~(\ref{eq:SO21_algebra}) and (\ref{eq:time_dependent_transformation}), Eq.~(\ref{eq:transformed_eom}) can be reduced to:

\begin{equation}
\partial_{t'}  \psi_s({\bf r}',t') = i \left[H_s, \psi_s({\bf r}',t')\right],
\end{equation}

\noindent where $H_s$ is defined in terms of the field operator, $\psi_s({\bf r}',t')$, and ${\bf r}'$ and $t'$ are the transformed coordinates given by either Eqs.~(\ref{eq:scale_inv}) or (\ref{eq:conf_inv}). Therefore the Heisenberg equation of motion for quantum systems with scale invariant Hamiltonians possesses both scale and conformal symmetry. This is the meaning of Fig.~(\ref{fig:geometry}), the Heisenberg equation of motion is equivalent in both the original and transformed space-time geometries. For a full proof of the invariance of the equation of motion under scale and conformal transformations, see Appendix \ref{app:eom}.

For the remainder of our discussions, we choose to work with a representation of the so(2,1) algebra in terms of differential operators. This can be done by noting that the infinitesimal generators in differential form can be defined as:

\begin{equation}
U_{\alpha}(\lambda) \psi_s({\bf r}, t) U^\dagger_{\alpha}(\lambda)|_{\lambda \rightarrow 0} = \left(1+\lambda G_{\alpha}\right)\psi_s({\bf r},t)+...
\end{equation}

\noindent where ${\alpha} = D, H_s, C$. The result is:

\begin{align}
G_{H_s} &= \partial_t,  \nonumber \\
G_{D}[\partial_{\bf r}, \partial_t] &= 2t \partial_t + \frac{d}{2} + {\bf r} \cdot \partial_{\bf r}, \nonumber \\
G_C^{\pm}[\partial_{\bf r}, \partial_t] &= t^2 \partial_t + t \left( \frac{d}{2} + {\bf r} \cdot \partial_{\bf r} \right) \pm i \frac{{\bf r}^2}{2}.
\label{eq:differential_so21_algebra}
\end{align}

\noindent Note that the $\pm$ in the generator of conformal transformations denote how the annihilation operator, $\psi_s({\bf r},t)$ (with negative "charge"), and the creation operator, $\psi_s^{\dagger}({\bf r},t)$ (with positive "charge"), transform. As can be readily checked, the operators, $iG_{H_s}$, $i G_C^{\pm}$ and $i G_D$ form a representation of the so(2,1) algebra, and satisfy Eq.~(\ref{eq:SO21_algebra}).

In this article, there are two important consequences of the SO(2,1) symmetry that we will exploit. The first concerns the spectrum of the oscillator Hamiltonian: $H_s + \omega^2 C$, where we will set $\omega$ to be unity for the remainder of the discussions. This Hamiltonian describes strongly interacting particles further confined in a harmonic potential. As reviewed in Appendix \ref{app:conformal_towers}, the spectrum of the oscillator Hamiltonian can be decomposed into a number of sets of states, where each state within a set is equally spaced from the next by two harmonic oscillator units. Each set is called a conformal tower, and can be labelled by the total number of particles, and the angular momentum quantum number, $l$, (we will ignore the azimuthal quantum number, $m$). The conformal tower state, and energy are given by:

\begin{align}
(H_s + C) | O^l_n \rangle &= E_n^l |O^l_n \rangle & E_n^l &= 2n + E_0^{l}.
\end{align}

\noindent The ground state energy of a given conformal tower, $E_0^{l}$, depends on the scaling dimension of a primary operator \cite{Wingate06, Nishida07}. These primary operators are discussed further in Appendix \ref{app:conformal_towers}.

The second feature of the SO(2,1) symmetry is concerned with the dynamics of these conformal tower states. Consider how a conformal tower state, $|O^{l}_n \rangle$, with energy $E_n^l$, evolves under the unitary evolution of a scale invariant Hamiltonian:

\begin{equation}
| O_n^l \rangle(t) = e^{-i H_s t} | O_n^l \rangle. 
\end{equation}

\noindent One can show using the so(2,1) algebra that these time evolved conformal tower states are the instantaneous eigenstates of a time dependent harmonic oscillator:

\begin{equation}
E_n^l |O_n^l\rangle(t) = (\tilde{H}_s + \tilde{C}) |O_n^l\rangle(t),
\end{equation}

\noindent where $\tilde{H}_s + \tilde{C}$ is given by:

\begin{align}
\tilde{H}_s + \tilde{C} &= (1+t^2) [ H_s ({\bf p} -\frac{{\bf r} t}{1+t^2}, {\bf r}) + \frac{C}{(1+t^2)^2}],
\label{instant}
\end{align}

\noindent and we have explicitly illustrated the scale invariant Hamiltonian's dependence on the momentum operator, ${\bf p}$, and position operator, ${\bf r}$.

The result here indicates that at arbitrary time, $t$, the time evolved conformal tower state, $|O_n^l\rangle(t)$ will be an eigenstate of the instantaneous Hamiltonian $H_I(t)=H_s({\bf p},{\bf r}) +C({\bf r})[1+t^2]^{-2}$, up to an overall gauge, with the eigenvalue equal to $E_n^l/(1+t^2)$. Alternatively, one can define a set of {\em generalized} conformal coordinates, $\tilde{\bf p}, \tilde{\bf r}$, such that:

\begin{align}
& \tilde{H}_s + \tilde{C} = H_s(\tilde{\bf p}, \tilde{\bf r}) +C(\tilde{\bf r}) & \nonumber\\
\tilde{{\bf p}} &= \sqrt{1+t^2}({\bf p} - \frac{{\bf r} t}{1+t^2}) & \tilde{{\bf r}} &= \frac{{\bf r}}{\sqrt{1+t^2}}. \nonumber \\
\label{eq:coordinates}
\end{align}

\noindent The full derivation of Eq.~(\ref{eq:coordinates}) is given in Appendix \ref{app:conformal_tower_dynamics}.

Eqs.~(\ref{instant}) and (\ref{eq:coordinates}) state two important intimately connected aspects of conformal tower state dynamics. One is that the dynamics of a conformal tower state is completely confined to the {\em eigenstate subspace} defined by the instantaneous Hamiltonian above. That is the instantaneous Hamiltonian, $H_I (t)$, effectively projects out a trajectory or path in the Hilbert space along which the many-body unitary evolution occurs. The second feature is that there are a set of convenient {\em generalized} conformal coordinates where the instantaneous Hamiltonian appears to be time independent and static. This possibility is directly and closely related to the space-time transformation described in the introduction, and in Fig.~(\ref{fig:geometry}).

Consequently, the many-body state, up to a gauge transformation, maintains its spatial profile in the generalized conformal coordinates, and its dynamics is  fully characterized by a space-time coordinate reparametrization (see below). This result can also be obtained using a wave function method, or by gauge potentials \cite{Castin04, Castin06,Gritsev10, Polkovnikov17}. This is the main reason to study the dynamics using the conformal tower basis. The conformal towers naturally encode the most elementary time dependent dynamics, allowing one to separate and focus on other more subtle dynamics such as interference effects.

\section{Density matrices and Conformal tower states}
\label{sec:den_matrix}

In this article, we will develop a microscopic approach to understand the dynamics of scale invariant and nearly scale invariant systems, by employing the $m$-body density matrix:

\begin{align}
P_m(\lbrace {\bf r}_{1i} \rbrace, &  \lbrace {\bf r}_{2i} \rbrace, t) =
\nonumber \\
& \langle \psi_0 | e^{i H t} \prod_{i=1}^m \psi^{\dagger}({\bf r}_{2i}) \prod_{i=1}^m \psi({\bf r}_{1i}) e^{-i H t} | \psi_0 \rangle.
\end{align}

\noindent Here we note that $|\psi_0 \rangle$ is the initial state, and for nearly scale invariant systems, $H$, the total Hamiltonian of the system, can be expanded around the scale invariant Hamiltonian: $H = H_s + V$, for some perturbation, $V$.

We restrict ourselves to a physical situation where the initial $N$-particle state is spatially localized so that the state can be 
conveniently expanded in terms of local conformal tower states $|O^l_n \rangle$, $n,l=0,1,2,...$ defined by the strongly interacting oscillator Hamiltonian $H_s+C$. This can be achieved in cold gases via laser confinement; again the harmonic frequency has been set to one for simplicity. 

The main observation one can make using the conformal tower states is that the dynamics of the density matrix has a highly generic form:

\begin{equation}
P_m(\lbrace {\bf r}_{1i} \rbrace, \lbrace {\bf r}_{2i} \rbrace, t) = \sum_{n,l; n',l'} \rho_{n,n'}^{l,l'}(\lbrace {\bf r}_{1i} \rbrace, \lbrace {\bf r}_{2i} \rbrace, t) \Gamma_{n',n}^{l',l}(t).
\label{eq:density_matrix_general}
\end{equation}

\noindent where:

\begin{align}
\rho_{n,n'}^{l,l'}(\lbrace {\bf r}_{1i} \rbrace, \lbrace {\bf r}_{2i} \rbrace, t) &= \langle O_n^l | \prod_{i=1}^m \psi_s^{\dagger}({\bf r}_{2i},t) \prod_{i=1}^m \psi_s({\bf r}_{1i},t) | O_{n'}^{l'} \rangle, \nonumber \\
\Gamma_{n',n}^{l',l}(t) &= \langle O^{l'}_{n'} | U_I(t) | \psi_0 \rangle \langle \psi_0 | U_I^{\dagger}(t) | O^l_n \rangle.
\label{eq:defs_matrices}
\end{align}

\noindent The unitary evolution operator, $U_I(t)$, is defined as:

\begin{equation}
U_I(t) = e^{i H_s t} e^{-i H t} = T \exp\left( -i \int_0^t dt' \ V_I(t') \right),
\label{eq:unitary_evolution_operator}
\end{equation}

\noindent with $T$ being the time ordering operator, and:

\begin{equation}
V_I(t) = e^{i H_s t} V e^{-i H_s t}.
\end{equation}

Eq.~(\ref{eq:density_matrix_general}) effectively separates the conformal symmetric dynamics from other contributions which potentially break conformal symmetries. The conformal invariant contribution is encoded in $\rho_{n,n'}^{l,l'}(\lbrace {\bf r}_{1i} \rbrace, \lbrace {\bf r}_{2i} \rbrace, t)$ which are defined in terms of the conformal tower states, and are independent of the initialization at $t=0$ (which usually break all symmetries) and $V$, the deviation from the fixed point Hamiltonian, $H_s$. The effects of symmetry breaking, either due to the initial conditions or due to the symmetry breaking interactions, are addressed by $\Gamma_{n',n}^{l',l}(t)$. In the next two sections, we will examine the scale invariant dynamics, and the consequences of the explicit symmetry breaking terms, respectively.

\section{Re-emergence of Conformal Symmetry in  Long Time Quench Dynamics}
\label{sec:re_emergence}

In the previous section we showed that it is possible to separate the dynamics governed by the scale invariant Hamiltonian, from the explicit symmetry breaking terms. 
In this section we will focus on the dynamics exactly at the fixed point, $H=H_s$ with $V=0$, and investigate the density matrix dynamics using the approach outlined in the previous section.

First, we will illustrate the general structure of the density matrix $\rho_{n,n'}^{l,l'}$ defined explicitly in terms of conformal tower states   

\begin{align}
\rho_{n,n'}^{l,l'}(\lbrace {\bf r}_{1i} \rbrace, \lbrace {\bf r}_{2i} \rbrace, t) = \langle O_n^l | \prod_{i=1}^m \psi_s^{\dagger}({\bf r}_{2i},t) \prod_{i=1}^m \psi_s({\bf r}_{1i},t) | O_{n'}^{l'} \rangle,
\end{align}

\noindent and how the SO(2,1) symmetry affects the dynamics.

As discussed in Appendix \ref{app:density_matrix}, it is possible to use the differential representation of the so(2,1) algebra, Eq.~(\ref{eq:differential_so21_algebra}), to obtain a first order partial differential equation for $\rho_{n,n'}^{l,l'}(\lbrace {\bf r}_{1i} \rbrace, \lbrace {\bf r}_{2i} \rbrace, t)$:

\begin{align}
0 &= \left[(1+t^2)\partial_t + t\sum_{i=1}^m \left( {\bf r}_{1i} \cdot \partial {\bf r}_{1i} + {\bf r}_{2i} \cdot \partial {\bf r}_{2i}+d \right)  \right. \nonumber \\
&+ \left. i \sum_{i=1}^m \frac{{\bf r}_{2i}'^2-{\bf r}_{1i}^2}{2} -i(E_n^l-E_{n'}^{l'})\right] \rho_{n,n'}^{l,l'}(\lbrace {\bf r}_{1i} \rbrace, \lbrace {\bf r}_{2i} \rbrace, t). \nonumber \\
\label{eq:diff_rho_s}
\end{align}

Substituting Eq.~(\ref{eq:diff_rho_s}) into  Eq.~(\ref{eq:density_matrix_general}) and taking the long time limit, one can show that the $m$-body density matrix satisfies the following differential equation:

\begin{align}
0 &= \left[\sum_{i=1}^m \left( G^+_C[\partial_{{\bf r}_{2i}},\partial_t] +G^-_C[\partial_{{\bf r}_{1i}}, \partial_t]\right) -(2m-1)t^2\partial_t \right] \nonumber \\
&P_m(\lbrace {\bf r}_{1i} \rbrace, \lbrace {\bf r}_{2i} \rbrace,t), \nonumber \\
\label{eq:diff_rho_s_3}
\end{align}

\noindent where $G_C^{\pm}$ is the generator of conformal transformations, defined in Eq.~(\ref{eq:differential_so21_algebra}).  Note that we are considering the fixed point Hamiltonian, $H_s$, with $V=0$.  The $\Gamma$ matrix is time independent, which leads to the above long time dynamics. 

Eq.~(\ref{eq:diff_rho_s_3}) states that at long times, the $m$-body density matrix for scale invariant fixed point interactions is an eigenfunction of the generator of conformal transformations, with zero eigenvalue. Therefore, the $m$-body density matrix must be left invariant under conformal transformations in the long time limit. Equivalently, one can show that the $m$-body density matrix must satisfy:

\begin{align}
P_m(\lbrace {\bf r}_{1i} \rbrace, \lbrace {\bf r}_{2i} \rbrace,t) &= \frac{1}{\left(1-\lambda t\right)^{dm}}e^{i \frac{1}{2}\sum_i (r_{2i}^2 - r_{1i}^2) \frac{\lambda}{1-\lambda t}} \nonumber \\
&P_m \left( \lbrace \frac{{\bf r}_{1i}}{1-\lambda t} \rbrace,\lbrace \frac{{\bf r}_{2i}}{1-\lambda t} \rbrace, \frac{t}{1-\lambda t} \right),
\label{cs}
\end{align}

\noindent in the long time limit, for arbitrary $\lambda$.

This is an intriguing result, as one would naively expect the scale symmetry to be present in the long time limit, along with conformal symmetry. However, this calculation shows explicitly that only conformal symmetry re-emerges in the long time limit. In addition, since the generators of conformal and scale transformations do not commute, it is generally impossible for a non-trivial density matrix to exhibit both symmetries in the long time dynamics. One can show that unless the density matrix is time independent, or there is no outward mass flow in space, scale and conformal symmetries are mutually exclusive. As we will discuss below, the re-emergence of conformal symmetry in Eq.~(\ref{cs}) will restrict the possible dynamics of the system, appreciably.

\section{The Action of Breaking Scale Symmetry}
\label{sec:breaksi}

In the previous section we focused primarily on the role of the SO(2,1) symmetry on the dynamics of systems with scale invariant interactions, and the emergence of conformal symmetry in the long time limit. In this section, we extend our analysis to the vicinity of the fixed point Hamiltonian, and consider $H=H_s+V$. We will examine how the dynamics are modified when scale invariance is explicitly broken by the interaction Hamiltonian.  This is tantamount to analyzing the matrix $\Gamma_{n',n}^{l',l}(t)$, defined in Eq.~(\ref{eq:unitary_evolution_operator}). The symmetry breaking perturbation $V$ was also analyzed previously in Ref.~\cite{Maki18}. Here we briefly summarize the results on $V$, and focus on the dynamics of a $m$-body density matrix.

We begin by examining how the perturbation, $V$, changes under a scale transformation. For our effective model, Eq.~(\ref{eq:Hamiltonian}), the perturbation can be shown to be proportional to:

\begin{align}
V &= \left(g(\Lambda)-g^*(\Lambda) \right) \int d{\bf r} \ \psi^{\dagger}({\bf r})\psi^{\dagger}({\bf r})\psi({\bf r})\psi({\bf r}) \nonumber \\
&= \frac{1}{\xi^{-\beta'(\tilde{g}^*(\Lambda))}} \frac{1}{\Lambda^{d-2-\beta'(\tilde{g}^*(\Lambda))}} \nonumber \\
&  sign \left[g(\Lambda) - g^*(\Lambda) \right]\int d {\bf r} \ \psi^{\dagger}({\bf r})\psi^{\dagger}({\bf r})\psi({\bf r})\psi({\bf r})
\label{eq:V_def}
\end{align}

\noindent where $\xi$ is the correlation length. For cold gases the length scale $\xi$ can be substituted for the d-dimensional scattering length, which parametrizes the strength of the interactions. Using this result one can show that under a scale transformation, the perturbation transforms as:

\begin{equation}
V \rightarrow e^{-(2 + \beta'(\tilde{g}^*))\lambda}V
\label{eq:V_rescaling}
\end{equation}

\noindent Eq.~(\ref{eq:V_rescaling}) states that the scaling dimension of the perturbation is $2 + \beta'(\tilde{g}^*)$. For this reason we define the shift in the scaling dimension as:

\begin{equation}
\alpha = -\beta'(\tilde{g}^*).
\end{equation}

\begin{figure}
\includegraphics[scale=0.31]{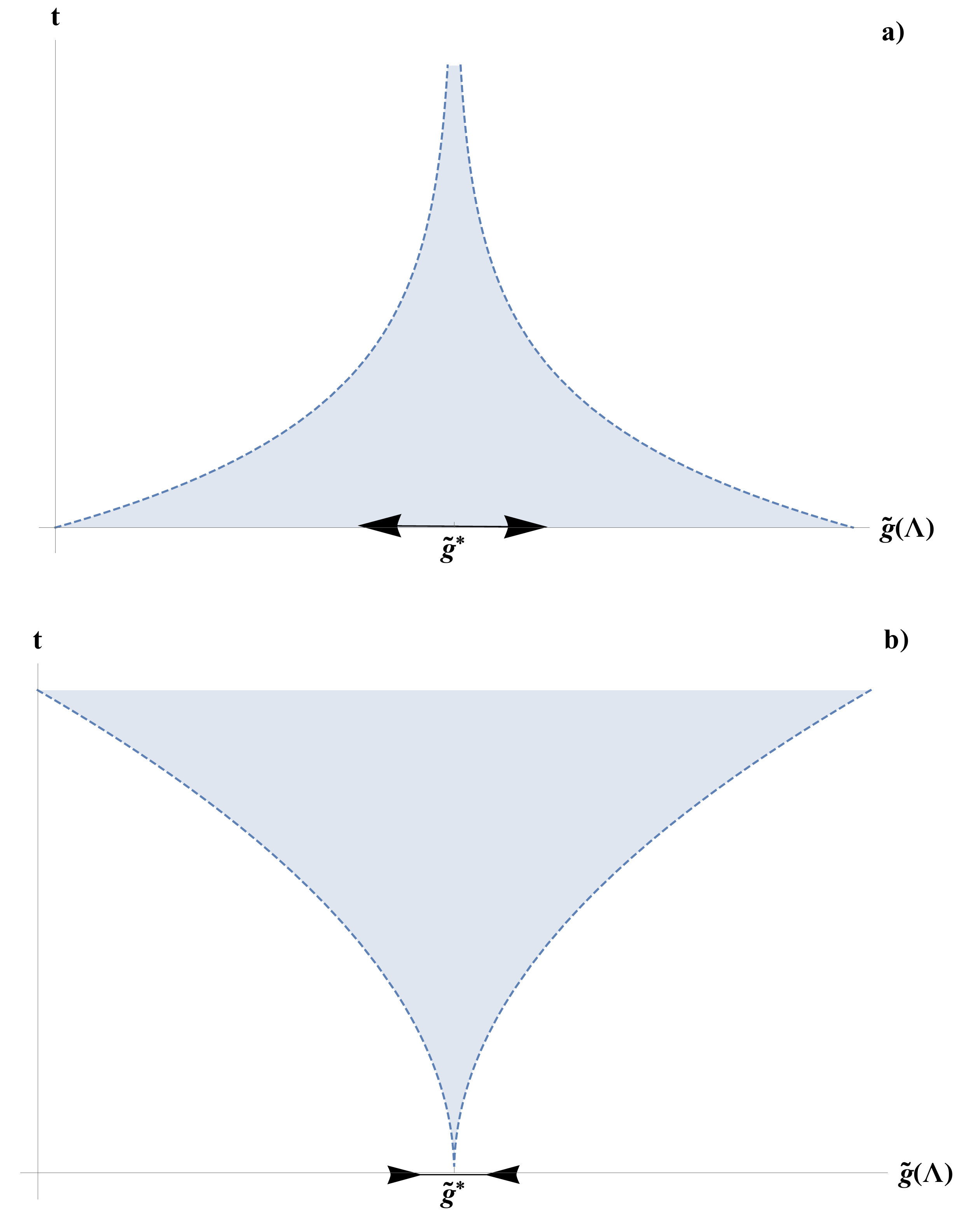}
\caption{Schematic of dynamics near an a) unstable and b) stable scale invariant fixed point, in the $t- \tilde{g}(\Lambda)$ plane. Here we focus on the asymptotic long time dynamics where $t \gg 1$, to avoid issues due to the initialization of the quantum state. In the shaded areas, the dynamics are governed by the emergent conformal symmetry. a) For an unstable fixed point, the region governed by the conformal symmetry is essentially the short time limit. For long times, the effect of the interaction becomes more dominating, and the emergent conformal symmetry will be broken,  which is represented by the non-shaded region. In this regime the dynamics will be modified by a non-trivial time dependence, which depends on how the perturbation changes under a scale transformation, see Eq.~(\ref{eq:V_def}). The arrows show the renormalization group flow out of the unstable fixed point. For three dimensional cold gases, this point corresponds to the resonant scale invariant fixed point. b) For a stable fixed point, any small perturbation will vanish in the long time limit. As a result, the long time dynamics will be governed by the emergent conformal symmetry. For three dimensional atomic systems, this point corresponds to the non-interacting limit.}
\label{fig:coupling_1}
\end{figure}

For perturbations with $\alpha < 1$, the effect of the perturbation will vanish in the long time limit. As a result, the dynamics are controlled by the scale invariant fixed point, $\tilde{g}^*$. In the case $\alpha \geq 1$, the interactions become relevant in the long time limit, and perturbation theory will break down after a certain time, and a non-perturbative solution is needed. Both these situations are explicitly shown in Fig.~(\ref{fig:coupling_1}). The shaded regime is where perturbation theory and the conformal dynamics are valid.

For the remainder of this article we will focus on unstable fixed points with $\alpha \geq 1$, as the dynamics will be modified by a non-trivial time dependence, in comparison to when the Hamiltonian has scale invariance. In this case, a non-perturbative solution for the unitary evolution operator exists in the long time limit:

\begin{equation}
U_I(t\gg 1) = \exp\left(-i \frac{t^{\alpha-1}}{\alpha-1} \frac{1}{\xi^{\alpha}} \tilde{V} \right),
\label{eq:U_I_longtime}
\end{equation}

\noindent where:

\begin{align}
\tilde{V} &= \Lambda^{2-\alpha - d} \int_{\Lambda} d {\bf r} \ \psi^{\dagger}(\bf{r})\psi^{\dagger}(\bf{r})\psi(\bf{r})\psi(\bf{r}), \nonumber \\
\tilde{V}_{n',l' ; n,l}&= e^{-i (E_n^l- E_n'^{l'}) \pi/2} \delta_{l,l'} \langle O_{n'}^{l'} | V | O_{n}^l \rangle.
\label{eq:V_tilde}
\end{align}

\noindent Note when $\alpha=1$, $U_I(t \gg 1)= \exp[-i\frac{\ln t}{\xi} \tilde{V}]$.

Eq.~(\ref{eq:U_I_longtime}) implies that in the long time limit the  matrix $\Gamma(t)$, defined in Eq.~(\ref{eq:defs_matrices}), satisfies:

\begin{equation}
\frac{\partial}{\partial t^{\alpha-1}/ \xi^{\alpha}} \Gamma(t) = -i \left[ \tilde{V}, \Gamma(t) \right].
\label{eq:gamma_eom}
\end{equation}

\noindent Since the derivative in the left hand side of Eq.~(\ref{eq:gamma_eom}) is taken with respect to $t^{\alpha-1}/\xi^{\alpha}$, it stands to reason that:

\begin{equation}
\Gamma(t) = \Gamma \left( \frac{t^{\alpha-1}}{\xi^{\alpha}} \right).
\end{equation}

\noindent This feature is robust and depends only on the renormalization group flow, and the universal scaling property of the perturbation operator, $V$, as it depends only on the basic symmetries and the dimensionality of the system. The microscopic details of the dynamics are encoded in the universal, dimensionless and regularized matrix, $\tilde{V}$, defined  in Eq.~(\ref{eq:V_tilde}). This matrix is the natural extension of the thermodynamic contact to a contact matrix \cite{Tan08, Zhang09}. 

The fact that the SO(2,1) symmetry is explicitly broken means a source term will appear in Eq.~(\ref{eq:diff_rho_s_3}). The modified differential equation for the density matrix in the long time limit is:

\begin{align}
&\left[\sum_{i=1}^m \left( G^+_C[\partial_{{\bf r}_{2i}},\partial_t] +G^-_C[\partial_{{\bf r}_{2i}}, \partial_t]\right) -(2m-1)t^2\partial_t \right] \nonumber \\
&P_m(\lbrace {\bf r}_{1i} \rbrace, \lbrace {\bf r}_{2i} \rbrace,t) \nonumber \\
&= \frac{t(\alpha-1)}{\xi^{\alpha}} \frac{\partial}{\partial \xi^{-\alpha}} P_m(\lbrace {\bf r}_{1i} \rbrace, \lbrace {\bf r}_{2i} \rbrace,t).
\label{eq:diff_rho_s_4}
\end{align}

\noindent Note for the case $\alpha = 1$, the source term is proportional to $t/(\xi \ln(t))$. For a detailed derivation of Eq.~(\ref{eq:diff_rho_s_4}),  see Appendix \ref{app:density_matrix}.

\begin{figure}
\includegraphics[scale=0.3]{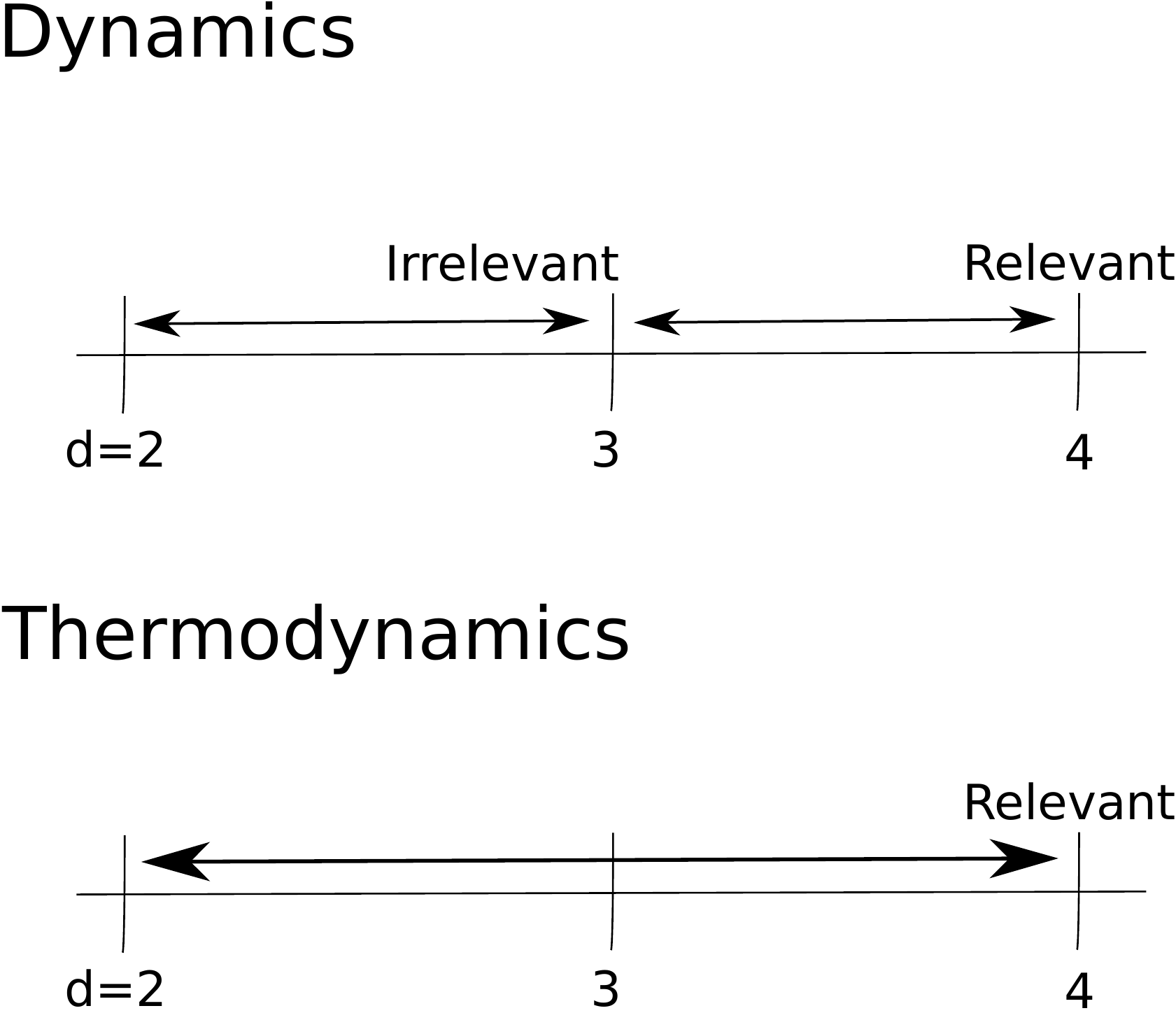}
\caption{Relevancy of perturbations to the resonant fixed point, valid for $2<d<4$. The top line gives the condition for relevancy to the dynamics, while the bottom line is for thermodynamics. Perturbations to the strong coupling fixed point are relevant for $d>2$ for thermodynamics, and $d>3$ for dynamics. The reason for the discrepancy is that the thermodynamics are governed by scale symmetry, while the dynamics are governed by conformal symmetry.}
\label{fig:relevancy}
\end{figure}

It is important to note that the condition for a relevant perturbation in thermodynamics is distinctly different in comparison to dynamics. A perturbation is relevant to the thermodynamics if $\alpha \geq 0$, and to the dynamics if $\alpha \geq 1$. This is because the thermodynamics and dynamics are governed by two different symmetries. In thermodynamics, it is the scale invariance, or scale symmetry, which defines the universal equation of states \cite{Ho04}. One only needs to consider how a perturbation rescales under a scale transformation to ascertain its relevancy, which is usually done via a standard scaling analysis. For quantum dynamics, it is conformal symmetry which defines the large scale - long time properties. This difference leads to the shift of the critical scaling dimension of the perturbation. In Fig.~(\ref{fig:relevancy}), we compare the relevancy of a perturbation to the resonant fixed point to the thermodynamics and dynamics of a d-dimensional atomic gas. As one can see, for dimensions $\leq 2 d \leq 4$ the perturbation to a strong coupling fixed point discussed above is relevant to the thermodynamics. As a result, the fixed point we have associated with $\tilde{g}^*$ is infra-red unstable (for the case $d=2$, the perturbation is marginally relevant). However for dynamics, the perturbation becomes relevant when $d\geq3$, with $d=3$ as the critical dimension where the perturbations are dynamically marginally relevant. For $3> d >2$, the perturbation, $V$, is thermodynamically relevant but dynamically irrelevant. 

Around the free fermion fixed point, with $\tilde{g}^*=0$, the situation is different. The free theory in $d=3$ is an infra-red stable fixed point, so that any perturbation around it is both thermodynamically and dynamically irrelevant. That is, deviations from scale invariance will not affect the long time emergent conformal symmetry, even if the Hamiltonian weakly break the scale symmetry (see discussions in Fig.~(\ref{fig:coupling_1})).

The same analysis can be carried out below two spatial dimensions, where there is a strong coupling fixed point at $\tilde{g}=2-d (>0)$. In the case $d=1$, all deviations, or perturbations, to this fixed point are irrelevant, i.e the strongly interacting fixed point is both  thermodynamically and dynamically infra-red stable. From this point of view, conformal dynamics in one spatial dimension are more robust than the three dimensional dynamics which we focus on in this article.

This section concludes our formal discussions of the density matrix. When the system has a scale invariant Hamiltonian, we exploited the consequences of the SO(2,1) - conformal symmetry, and its resultant conformal tower states on the dynamics of the density matrix. We have also classified dynamically relevant and irrelevant perturbations which break the scale symmetry explicitly. The result of this analysis is a partial differential equation describing the dynamics of the density matrix: Eq.~(\ref{eq:diff_rho_s_4}). In the following discussions, we will use the $m$-body density matrix to investigate the manifestations of the emergent conformal symmetry, and its breaking, on the dynamics of many body systems.

\section{Signatures of conformal towers in hydrodynamic flows}
\label{sec:hydro_flow}


We begin by discussing the emergent conformal symmetry and its breaking on the hydrodynamic flows of the three dimensional unitary Fermi gas. We consider a three dimensional Fermi gas initially placed in a harmonic trap, either isotropic or anisotropic. At $t=0$, the trap is released, while the resonant interactions are maintained. This was done experimentally in Refs.~\cite{O'Hara02,Kinast04,Kinast06,Cao11,Elliott14} and analysed using a variational solution for the hydrodynamic equations of motion. Although the hydrodynamic approach can accurately describe the experiment, it is a phenomenological approach that does not explicitly highlight the effect of the emergent conformal symmetry and its breaking. 

\begin{figure}
\includegraphics[scale=0.25]{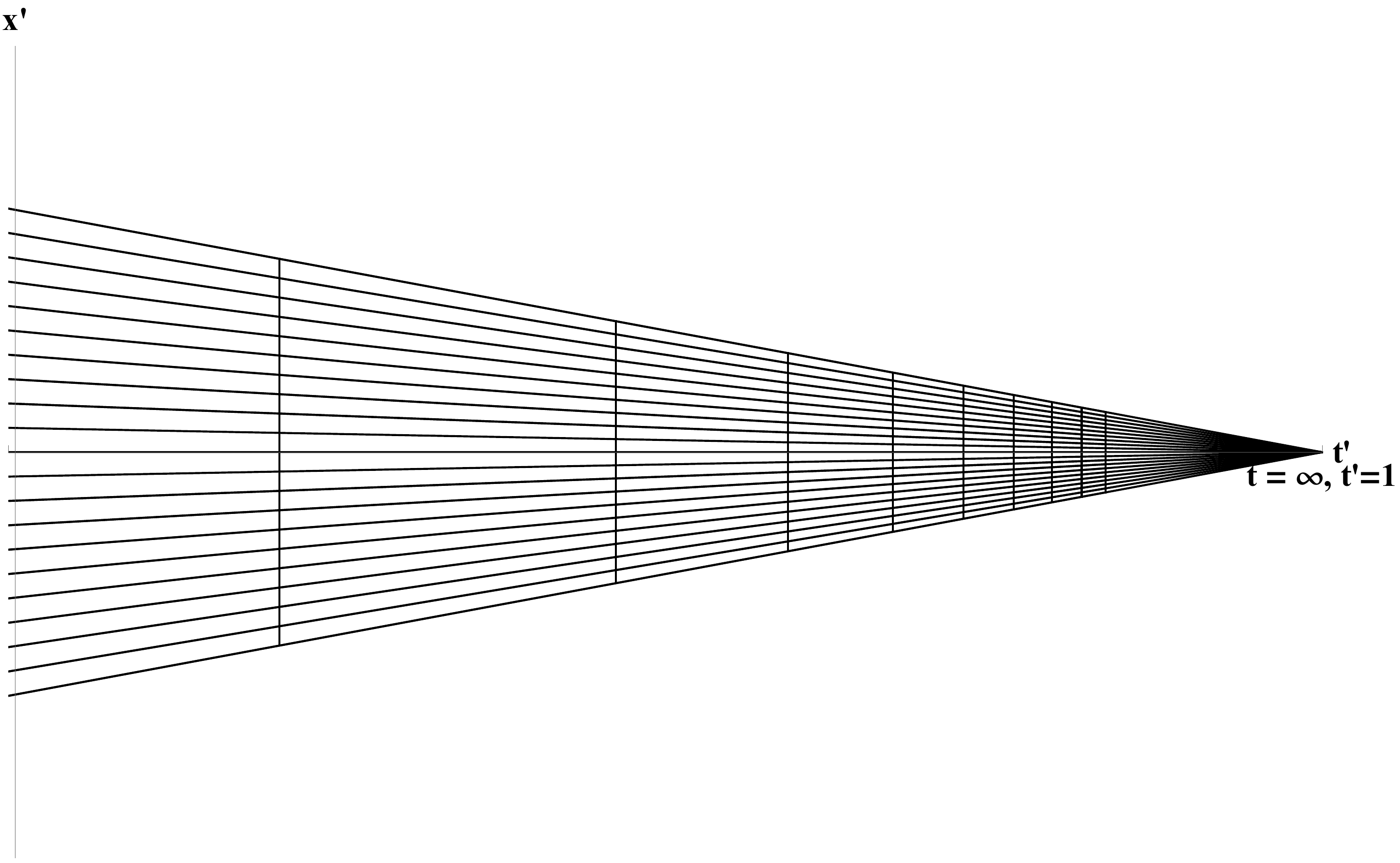}
\caption{Quench space-time geometry. Due to the emergent conformal symmetry, the physics in both the original, and the above space-time are equivalent. All the long time dynamics of local observables can be determined by determining the dynamics for $t' \approx 1$.}
\label{fig:quench_geometry} 
\end{figure}

In this section we will examine the expansion dynamics using the reduced one-body density matrix, as this approach will explicitly elucidate the role of the emergent conformal symmetry, and its breaking. As discussed in the introduction, this conformal symmetry states that the physics will be identical in a modified space-time geometry. For the quench experiment, this modified space-time geometry is shown in Fig.~(\ref{fig:quench_geometry}). One can again see that the long time dynamics can be understood by taking the limit $t' \rightarrow 1$, and then transform back to the original coordinates $({\bf r},t)$. We will exploit this fact to determine the asymptotic long time behaviour of the expansion dynamics.

In order to study the expansion dynamics, we introduce the moment of inertia tensor:

\begin{align}
I_{i,j}(t) &= \langle r_i r_j \rangle(t) = \int d {\bf r} \ r_i r_j P_1({\bf r},{\bf r},t) & i,j &= x,y,z,
\label{eq:moment_of_inertia_def}
\end{align}

\noindent where $P_1({\bf r}, {\bf r},t)$ is the one-body reduced density matrix. In Eq.~(\ref{eq:moment_of_inertia_def}), the positions of the $i$th particle are measured with respect to the center of mass of the gas. For an initial state which is time-reversal invariant, the center of mass coordinate will remain at the origin for all times. For the remainder of our discussions we will focus on quantum states which are initially time-reversal invariant.

In general, this tensor can be decomposed into two pieces which are labelled by how they transform under rotations:

\begin{align}
I_{i,j} &= \frac{1}{3} \langle r^2 \delta_{i,j} + Q_{i,j} \rangle(t) \nonumber \\
\langle Q_{i,j}\rangle(t) &= \int d{\bf r} \ (3 r_i r_j - r^2 \delta_{i,j}) P_1({\bf r},{\bf r},t)
\label{eq:moment_decomposition}
\end{align}

\noindent The first term in Eq.~(\ref{eq:moment_decomposition}), $\langle r^2 \rangle(t)$, is the monopole moment which is isotropic; i.e. it carries angular momentum, $l = 0$. The second piece, $\langle Q_{i,j}\rangle(t)$, is a traceless symmetric tensor, known as the quadrupole moment, which carries angular momentum, $l=2$. We will focus on both the monopole and quadrupole moments, and see how the conformal symmetry and its breaking affect the long time dynamics.

\subsection{Expansion Dynamics for the Scale Invariant Fermi Gas}

We begin by considering the expansion dynamics of a Fermi gas at resonance, i.e. $V=0$ or $\xi = \infty$. At time $t=0$, the gas is released from  the trap, and the dynamics are governed by a scale invariant Hamiltonian, $H_s$. We can then utilize the SO(2,1) symmetry to write down a differential equation for the moment of inertia tensor. To do this, we note that the moment of inertia depends on the one-body density matrix which is subject to Eq.~(\ref{eq:diff_rho_s_3}) in the long time limit. Utilizing the SO(2,1) symmetry, one can show that the moment of inertia tensor must satisfy:

\begin{align}
\left((1 \right. &+  \left. t^2) \partial_t - 2 t \right) I_{i,j}(t) = \nonumber \\ 
& \sum_{n,l; n', l'} i(E_n^l - E_{n'}^{l'}) \int d{\bf r} \ r_i r_j \rho_{n,n'}^{l,l'}({\bf r},{\bf r},t) \Gamma_{n',n}^{l',l}(0).
\label{eq:diff_moment_inertia}
\end{align}

The first line describes the long time limit and is the result of the emergent conformal symmetry, while the second describes the leading long time corrections. Taking the long time limit, one can evaluate Eq.~(\ref{eq:diff_moment_inertia}) exactly to obtain:

\begin{align}
I_{i,j}(t) &= \left(\frac{t}{t_0}\right)^2 I_{i,j}(t_0) & t,t_0 &\gg 1.
\label{eq:conformal_moment}
\end{align}

\noindent Therefore, conformal symmetry requires that the moment of inertia tensor be proportional to $t^2$ in the long time limit. In this limit, the dynamics are equivalent to a time dependent rescaling; understanding the moment of inertia at one point in time guarantees knowledge at future times.


The leading correction to the moment of inertia tensor will depend on the initial conditions, $\Gamma(0)$. In particular, there are two main cases one can consider: the expansion from an isotropic trap, and from an anisotropic one.

\subsubsection{Leading Long Time Correction for Isotropic Expansion Flow}

For expansions from an isotropic trap, the initial Hamiltonian possesses SO(2,1) symmetry, and as a result rotational symmetry. Therefore, angular momentum is a good quantum number throughout the whole expansion. We will focus on systems that are initially isotropic,  so that we only need to consider the coupling between states in the s-wave conformal tower. In this case, the quadrupole moment will vanish for all times, and one only needs to focus on the monopole moment.

There are two generic types of initial conditions one can prepare for an isotropic trap. The first is a unitary Fermi gas prepared in equilibrium, or equivalently in an exact conformal tower state. In these cases the matrix, $\Gamma(0)$, will be diagonal. In particular, for thermal equilibrium, the initial conditions have the form: $\Gamma(0)_{n',n}^{l',l} = \delta_{0,l} \delta_{l,l'} \delta_{n,n'} e^{- E_n^l/T_0}$, where $T_0$ is the initial temperature of the system. In this case, there is no interference between the conformal tower states, and the solution for the monopole moment is:

\begin{equation}
\langle r^2 \rangle(t) = \langle r^2 \rangle(0) \left( \omega^2 t^2+1\right).
\label{eq:monopole_diagonal}
\end{equation}

\noindent Eq.~(\ref{eq:monopole_diagonal}) is exact at all times. 

In the long time limit $\langle r^2 \rangle(t) \propto t^2$ as required by conformal symmetry. We define the coefficient of proportionality to be $v^2$, where we call $v$ the relative velocity:

\begin{equation}
v = \lim_{t \rightarrow \infty}\sqrt{\frac{\langle r^2 \rangle(t)}{\langle r^2 \rangle(0) t^2}},
\end{equation}

\noindent For the case of a system initially placed in thermal equilibrium, the relative velocity is pinned to the trap frequency. This is a consequence of the Feynman-Helmann theorem, or equivalently, the SO(2,1) symmetry. We finally note that all information of the initial conditions, such as initial temperature and the Fermi energy, are contained in $\langle r^2 \rangle(0)$.

A second more general class of initial conditions can be prepared by taking a non- or weakly-interacting Fermi gas at temperature, $T_0$, and quenching it to unitarity. In this case, the initial conditions will be a non-trivial superposition of resonant s-wave conformal tower states.  In this case, there will be interference from states within the s-wave conformal tower, but each conformal tower will be decoupled from another. In this case, one can show that the long time dynamics follow:

\begin{equation}
\langle r^2 \rangle(t \rightarrow \infty) \approx \left(v^2 t^2 + B\right) \langle r^2 \rangle(0).
\label{eq:monopole_dynamics_general}
\end{equation}

\noindent where $B$ is a constant that depends on the interference between states within a single conformal tower. For an explicit expression, see Appendix \ref{app:flow}. The relative velocity will also depend on the interference of the conformal tower states. In particular, one can show using the Heisenberg equations of motion for the monopole moment that the relative velocity can be written as:

\begin{equation}
v = \sqrt{\frac{2 \langle H_s \rangle}{\langle r^2 \rangle(0)}}.
\end{equation}

\subsubsection{Leading Long Time Correction for Elliptic Flow}

Consider an anisotropic trap, with frequencies $\omega_i$, with $i = x,y,z$. Since the initial trap is anisotropic, the SO(2,1) symmetry is initially broken. As a result, the initial state will be projected into a number of different conformal towers, and it is impossible to prepare a diagonal ensemble in the conformal tower basis. In this case, the expansion dynamics will have non-vanishing monopole and quadrupole moments. The dynamics of the monopole moment will be identical to the isotropic case. In this section, we will focus on the dynamics of the quadrupole moment. 

The quadrupole moment will still satisfy Eq.~(\ref{eq:diff_moment_inertia}). In the long time limit, the quadrupole moment must be proportional to $\langle Q_{i,j} \rangle (t) \propto t^2$, due to the conformal symmetry. The relative velocities will differ for different directions, resulting in what is known as elliptic flow. The key difference between the isotropic compressional flow from the anisotropic elliptic flow is in the leading order correction to the conformal dynamics. To see this, note that:

\begin{equation}
\int d{\bf r} \ Q_{i,j} \ \rho_{n,n'}^{l,l'}({\bf r}, {\bf r},t) \propto \delta_{l,l'=l \pm 2}.
\label{eq:wig_eck}
\end{equation}

\noindent Eq.~(\ref{eq:wig_eck}) follows from Wigner-Eckhardt theorem, and states that the quadrupole moment will couple different conformal towers together: $l' = l \pm 2$. This is in contrast to the compressional flow, where the  monopole moment does not couple different conformal towers together. 

As seen in Appendix \ref{app:flow}, this inter-tower interference leads to a correction to the dynamics that is linear in time:

\begin{equation}
\langle Q_{i,j}\rangle(t \rightarrow \infty) \approx \left( v^2_{i,j} t^2 + A_{i,j} t + B_{i,j} \right),
\label{eq:quadrupole_dynamics}
\end{equation}

\noindent where  $v^2_{i,j}$, $A_{i,j}$ and $B_{i,j}$ are traceless, symmetric tensors. Note that  $A_{i,j}$ depends only on the inter-tower interference, and is unique to elliptic flow, while $B_{i,j}$ will depend on both inter- and intra-tower interference.

In the presence of azimuthal symmetry along the $z$-direction, we can further reduce Eq.~(\ref{eq:quadrupole_dynamics}) to:

\begin{align}
\langle Q_{i,j} \rangle(t) &= 3 Q(t) ({{\bf e}_z}_i{{\bf e}_z}_j-\frac{1}{3} \delta_{ij}) \nonumber\\
Q(t \rightarrow \infty) &\approx v^2_Q t^2 +A_Q t + B_Q.
\label{eq:Q_moment}
\end{align}
\noindent with constants $v^2_Q$, $A_Q$, and $B_Q$ depending on the initial conditions, and ${\bf e}_z$ is the unit vector along $z$-axis. This result describes elliptic flow of scale invariant interactions, elongated along the $z$ direction.

\subsection{Expansion Dynamics for Nearly Scale Invariant Systems}
The previous discussions were focused on the dynamics of the Fermi gas at resonance, when the interaction Hamiltonian is scale invariant. 
Here we turn to the vicinity of the fixed point, and turn on a small deviation, $V$. 
We assume that the scattering length is large but finite, $\infty > \xi \gg 1$.  For three-dimensional Fermi gases, one can show that $\alpha = 1$. Therefore, the scale invariant dynamics will be modified by a function of  $\ln(t)/\xi$. The dynamics for the monopole and quadrupole moments will have the form:

\begin{align}
\langle r^2 \rangle (t\rightarrow \infty) &\approx v^2\left( \frac{\ln(t)}{\xi}\right) t^2 + B\left( \frac{\ln(t)}{\xi}\right) \nonumber \\
\langle Q_{i,j} \rangle (t \rightarrow \infty) &\approx  v_{i,j}^2\left(\frac{\ln(t)}{\xi}\right) t^2 + A_{i,j}\left(\frac{\ln(t)}{\xi}\right)t \nonumber \\
 & + B_{i,j}\left(\frac{\ln(t)}{\xi}\right),
\end{align}

\noindent where all the constants (tensors) are now functions of $\ln(t)/\xi$.

In the presence of azimuthal symmetry, we can again reduce the quadrupole moment to a form equivalent to Eq.~(\ref{eq:Q_moment}). However, now the function, $Q(t)$, is modified by the breaking of scale invariance:

\begin{equation}
Q(t\rightarrow \infty) \approx  v_Q^2\left( \frac{\ln(t)}{\xi}\right) t^2 + A_Q\left( \frac{\ln(t)}{\xi}\right) t + B_Q\left( \frac{\ln(t)}{\xi}\right).
\end{equation}

\noindent As in the previous case, $v_Q^2$, $A_Q$, and $B_Q$, are now all functions of $\ln(t)/\xi$.

\subsection{Comparison to Variational Hydrodynamics}

In Ref.~\cite{Elliott14}, the expansion dynamics of the Fermi gas were studied for the resonant, and nearly resonant Fermi gas. In their work they prepared the Fermi gas in an anisotropic harmonic trap, and examined the elliptic flow. In a similar fashion, consider an azimuthally symmetric harmonic potential with frequencies $\omega_{\perp}$ in the $x-y$ plane, and $\omega_z$ in the $z$-direction. We parametrize the elliptic flow by examining the aspect ratio: the ratio of the moment of inertias in the $z$-direction, and a given direction in the $x-y$ plane, say the $x$-direction. Using Eq.~(\ref{eq:moment_decomposition}), one can see that:

\begin{eqnarray}
&& I_{i,j}(t) = \frac{1}{3} \langle r^2 \delta_{i,j} + Q_{i,j} \rangle(t); \nonumber \\
&& \langle r^2 \rangle(t) \approx \left(v^2 t^2 + B\right) \langle r^2 \rangle(0), \nonumber \\
&& \langle Q_{i,j} \rangle(t) = 3 Q(t) ({{\bf e}_z}_i{{\bf e}_z}_j-\frac{1}{3} \delta_{ij}) 
\end{eqnarray}

where  $Q(t) \approx v^2_Q t^2 +A_Q t + B_Q$ as defined before, and $i = x,y,z$. As seen in Eqs.~(\ref{eq:moment_decomposition}) and Eq.~(\ref{eq:Q_moment}), the moment of inertia is a diagonal matrix with two distinct eigenvalues  in the case of azimuthal symmetry, rather than three as in most generic cases. One can show that the asymptotic dynamics for the the aspect ratio will have the form:

\begin{align}
\frac{I_{x,x}(t)}{I_{z,z}(t)} &= \frac{\langle r^2 \rangle(t) -Q(t)}{\langle r^2 \rangle(t) + 2 Q(t)} \nonumber \\
&\approx \frac{v^2 \langle r^2 \rangle(0)- v_Q^2}{v^2 \langle r^2\rangle(0) + 2 v_Q^2} \nonumber \\
&- \left(\frac{A_{Q}}{v^2 \langle r^2\rangle(0) - v_Q^2} + \frac{2A_{Q}}{v^2 \langle r^2\rangle(0)+ 2v_Q^2} \right) \frac{1}{t}.
\label{eq:aspect_ratio}
\end{align}

\noindent As one can see, the aspect ratio must saturate to a finite value in the long time limit as a direct consequence of conformal symmetry. The aspect ratio approaches a constant value with a correction of order $O(t^{-1})$ which we associate with the inter-tower interference effects between conformal tower states.

In Ref.~\cite{Elliott14}, the authors performed a similar experiment, and examined the aspect ratio due to the anisotropy in the $x-y$ plane. They observed that the growth of the aspect ratio slowed down, but the saturation was not explicitly confirmed. However, the $t^2$ dependence of the monopole moment was confirmed. If the experiment was repeated, and the expansion dynamics were tracked for longer times, the saturation of the aspect ratio will be more visible; this is a signature of the emergent conformal symmetry in elliptic flow.

The early hydrodynamical approach \cite{Elliott14, Stringari02} and their results are consistent with our general density matrix approach. The consistency justifies various assumptions previously made in obtaining the variational solutions to the hydrodynamics equations. The density matrix approach here, on the other hand, illustrates explicitly the role of the re-emergent conformal symmetry, and the relationship between the leading long time correction and the interference between conformal tower states.

\section{Thermal Entropy in Asymptotic Dynamics}
\label{sec:ent}

\subsection{Conformal Symmetry and Entropy Conservation}
 
We now turn to another physical quantity, the thermal entropy. Let us consider a Fermi gas of $N$ particles initially localized in space. As a result, the initial state can be expanded in terms of conformal towers and we again use $\Gamma_{n,n'}^{l,l'}(0)$
represent a general mixture. The thermal entropy of a Fermi gas of $N$ particles is given by:

\begin{equation}
S(t) = - Tr \left[ P_N \log P_N \right],
\end{equation}

\noindent where $\log P_N$ is the logarithm of the $N$-body density matrix defined via:

\begin{equation}
P_N = e^{\log P_N} = \sum_{n=0}^{\infty} \frac{1}{n!} (\log P_N)^n.
\end{equation}

\noindent In addition, it is possible to define an entropy density:

\begin{align}
S({\bf r},t) &= -\int \prod_{i=1}^N d{\bf r}_{1i} \ d{\bf r}_{2i} \  \delta({\bf r}-{\bf r}_{1i=1}) \nonumber \\
&P_N(\lbrace {\bf r}_{1i} \rbrace, \lbrace {\bf r}_{2i} \rbrace,t) \log P_N (\lbrace {\bf r}_{2i} \rbrace, \lbrace {\bf r}_{1i} \rbrace,t)
\end{align}

We begin our discussion by considering the dynamics of the entropy density when the Hamiltonian is scale invariant. Using the SO(2,1) symmetry, one can show that the emergent conformal symmetry restricts the entropy density. In fact, as seen in Appendix \ref{app:entropy}, one can show that the entropy density satisfies the following continuity equation:

\begin{align}
\partial_t S({\bf r}, t) + \partial {\bf r} \cdot [\frac{t \bf r}{1+t^2} S({\bf r},t)]=0.
\label{eq:entropy:continuity}
\end{align}

\noindent In the long time limit, this differential equation reduces to:

\begin{align}
[ G^+_c[\partial_{\bf r}, \partial_t]+G_c^{-} [\partial_{\bf r}, \partial_t] - t^2 \partial_t] S({\bf r}, t)=0
\label{eq:entropy:continuityA}
\end{align}

\noindent which indicates the entropy density field shall be an eigenstate of the conformal generator with zero eigenvalue. That is the solution to this equation has to be conformal invariant and hence satisfies the following identity

\begin{align}
S({\bf r}, t)=\frac{1}{(1-\lambda t)^d} S(\frac{{\bf r}}{1-\lambda t},\frac{t}{1-\lambda t})
\end{align}
for arbitrary $\lambda$.

Eq.~(\ref{eq:entropy:continuity}) explicitly asserts that the entropy is overall conserved:

\begin{equation}
\partial_t S(t) = 0.
\end{equation}

\noindent This can be understood in two ways. The first was mentioned in the introduction. Conformal symmetry means that the dynamics of the system are equivalent in both space-time geometries represented in Fig.~(\ref{fig:geometry}). The long time dynamics of the entropy is equivalent to the dynamics near $t' \approx 1$ in the modified space-time geometry. Since the entropy is a dimensionless quantity, we expect the entropy to simply saturate. To see this clearly, we note that conformal symmetry implies:

\begin{equation}
S(t) = S\left(t'=\frac{t}{1- \lambda t}\right),
\end{equation}

\noindent for arbitrary $\lambda$. One can see that the entropy must be a constant function of $t$, and hence is conserved. The second more microscopic explanation is that during the expansion dynamics, the probability of the system being in a given dynamically evolving conformal tower state is conserved. As a result, the entropy must saturate to a finite value. In fact, the density matrix $\Gamma$, spanned over the conformal tower states, doesn't have any dynamics in the long time limit, when the Hamiltonian is scale invariant; so for a mixture, thermal or not, the entropy is always strictly conserved.
This microscopic consideration naturally leads to the concept of {\em conformal cooling}, which we will discuss in the next subsection.

\subsection{Conformal Cooling As a Result of Conformal Symmetry}

In this subsection, we will establish that the entropy conservation in the unitary time evolution of a scale invariant Hamiltonian naturally leads to the notion of conformal cooling, which distinctly differs from more conventional adiabatic cooling that usually involves losses of internal energy. We present our analysis for the case when the initial quantum state is an equilibrium state of $H_s+C$, with an initial temperature $T_0$. This can be achieved by confining a gas in a harmonic potential (where again we have set the harmonic frequency to be unity). For this initial state, the $\Gamma$ matrix is diagonal and time independent:
 
\begin{equation}
\Gamma_{n',n}^{l',l}=N\delta_{n,n'}\delta_{l,l'} \exp(-\frac{E_n^l}{\kappa T_0}),
\end{equation}
\noindent 

\noindent where $N$ is a normalization factor, and $\kappa$ is Boltzmann's constant.

At $t=0$, the trapping potential is turned off, and the gas expands in the presence of scale invariant interactions, i.e.  the unitary evolution is governed by the scale invariant Hamiltonian, $H_s$.

The dynamics of the $m$-body reduced density matrix, $P_m (t)$ are governed by the following equation, valid for arbitrary $t$:

\begin{align}
0 &= \left[(1+t^2)\partial_t + t\sum_{i=1}^m \left( {\bf r}_{1i} \cdot \partial {\bf r}_{1i} + {\bf r}_{2i} \cdot \partial {\bf r}_{2i}+d \right)  \right. \nonumber \\
&+ \left. i \sum_{i=1}^m \frac{{\bf r}_{2i}'^2-{\bf r}_{1i}^2}{2}  \right]
P_m(\lbrace {\bf r}_{1i} \rbrace, \lbrace {\bf r}_{2i} \rbrace, t).
\label{eq:diff_P}
\end{align}

\noindent In the long time limit, this differential equation again reduces to the form in Eq.(\ref{eq:diff_rho_s_3}).



For simplicity, we will consider the one-particle density matrix, which has the following form  suggested by the invariance of the density matrix under a conformal transformation, see Eq.~(\ref{eq:diff_P} ):

\begin{eqnarray}
&& P_{m=1}({\bf r}_1, {\bf r}_2; t)=\frac{1}{(1+t^2)^{d/2}} \exp\left[i \frac{r_1^2 -r_2^2}{2} \frac{t}{1+t^2}\right]  
\nonumber \\
&& P^{0}_{m=1}(\frac{{\bf r}_1}{(1+t^2)^{1/2}}, \frac{{\bf r}_2}{(1+t^2)^{1/2}})
\label{solution}
\end{eqnarray}

Now the initial density matrix is given by:

\begin{equation}
P^0_{m=1}({\bf r}_1, {\bf r_2}) = P^{eq}_{m=1}({\bf r}_1, {\bf r_2}; T_0).
\end{equation}

\noindent The reparameterization of the spatial coordinates in
Eq.~(\ref{solution}) suggests that at time $t$ the density matrix (up to the gauge factor) must maintain an equilibrium form with a rescaled harmonic frequency:

\begin{eqnarray}
&& P_{m=1}({\bf r}_1, {\bf r}_2; t)=...\nonumber \\
&& P^{eq}_{m=1}(\frac{{\bf r}_1}{(1+t^2)^{1/2}}, \frac{{\bf r}_2}{(1+t^2)^{1/2}}; \frac{T_0}{1+t^2}\cdot (1+t^2) )
\label{eq:conformal_cooling}
\end{eqnarray}

\noindent where we use $...$ to represent the gauge term which plays no role in the remainder of our discussion. Eq.~(\ref{eq:conformal_cooling})  is identical to an equilibrium density matrix for an instantaneous Hamiltonian, $H_I(t)=H_s+C (1+t^2)^{-2}$, at a rescaled temperature:

\begin{align}
T(t) &=\frac{T_0}{1+t^2} \rightarrow \frac{T_0}{t^2} & &\mbox{as $t\rightarrow \infty$}
\end{align}

\begin{figure}
\includegraphics[scale=0.4]{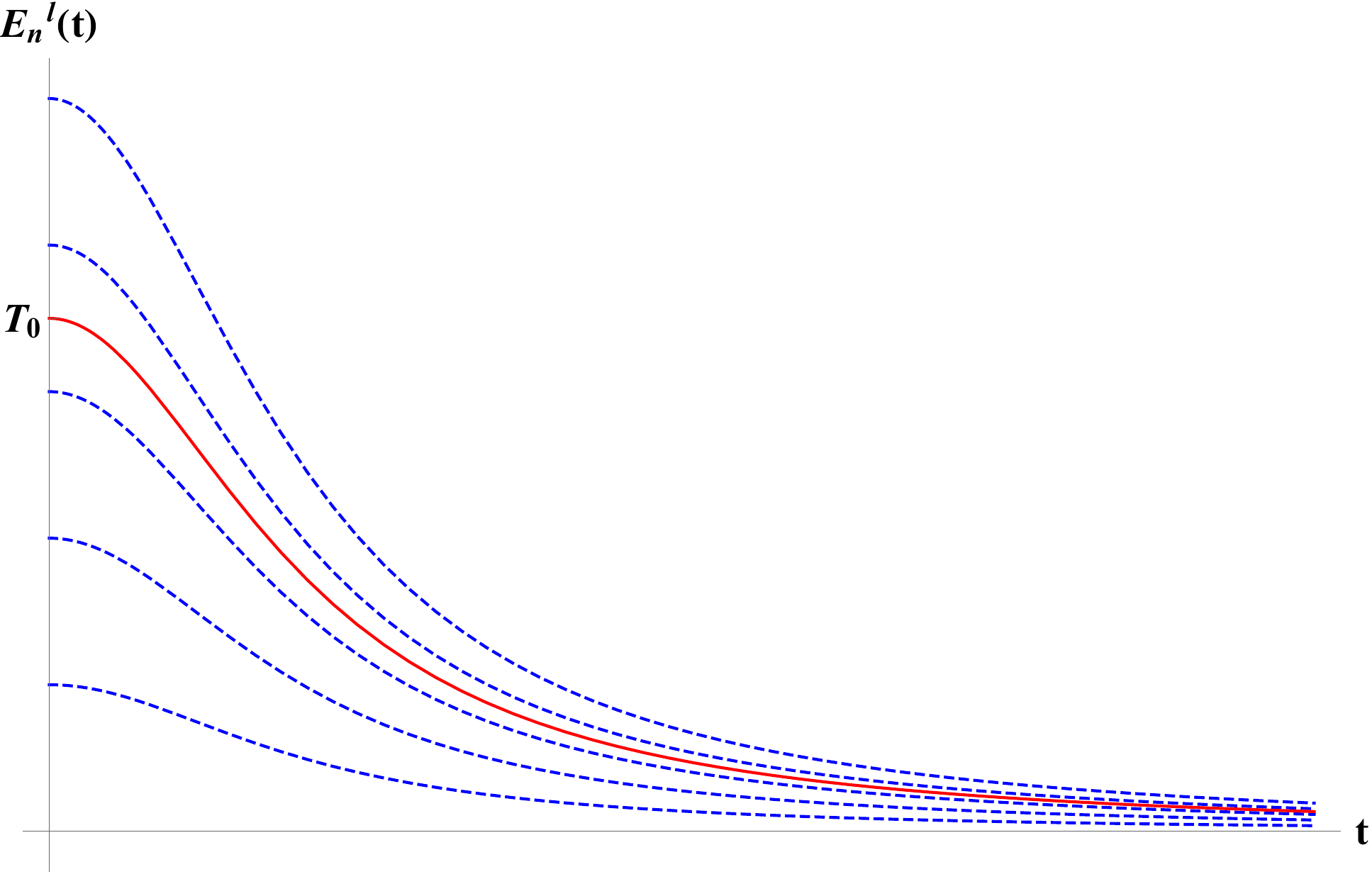}
\caption{The time dependence of the temperature compared to the time dependence of the eigenvalues of an instantaneous Hamiltonian: $H_I(t)$, see main text. The eigenvalues are given by: $E_n^l(t)=E_n^l/(1+t^2)$, shown in blue, while the temperature: $T(t)=T_0/(1+t^2)$, is shown in red. As a result, the Boltzmann weight for a given conformal tower state is a conserved quantity - leading to the conservation of entropy for scale invariant systems.}
\label{fig:conf_cooling}
\end{figure}

As a result, the entropy is conserved because the time dependence of the eigenvalues of the instantaneous Hamiltonian $H_I(t)$ (i.e. $E_n^l(t)=E_n^l/(1+t^2)$), is equivalent to the time dependence of the temperature. This is shown explicitly in Fig.~(\ref{fig:conf_cooling}). The fact that the temperature evolves like the instantaneous eigenvalues of $H_I(t)$ implies that the Boltzmann weight of the instantaneous conformal tower states associated with
$H_I(t)$, is an invariant:

\begin{eqnarray}
\exp(-\frac{E_n^l}{\kappa T_0})=\exp(-\frac{E_n^l (t)}{\kappa T(t)}), 
\end{eqnarray}

We denote this as {\em conformal cooling}, i.e. cooling in a pure statistical sense from a conformal-tower-state point of view. This cooling is due to the compression of the conformal tower spectra of the instantaneous, $H_I(t)$, in the energy space as $t$ increases. It is worth re-emphasizing that the interacting many-body system here is simply undergoing a unitary evolution, defined by the scale invariant Hamiltonian, $H_s$, without any other physical confining potentials. As was discussed previously, the dynamics of the system can be fully mapped onto the adiabatic dynamics characterized by the instantaneous Hamiltonian $H_I(t)$ -  although $H_I(t)$ involves a fictitious time dependent confining potential. The instantaneous conformal tower state basis provides the most convenient representation for dynamics. We refer the reader to Sec.~\ref{sec:scale_conf} for more discussions on $H_I(t)$ and the conformal tower states.

Entropy conservation, and consequentially, conformal cooling in a freely expanding interacting gas are highly surprising, given that the free expansion of a classical thermal gas (into a vacuum) always involves non-equilibrium states that result in entropy production, as we have been taught in elementary thermal physics.  On the other hand, it illustrates that in the quantum dynamics, scale invariant interactions enforce that at time $t$ the dynamical state is effectively an equilibrium state of an instantaneous Hamiltonian, $H_I(t)$. Hence, it is possible to introduce an equations of state to characterize the dynamics. This is what the zero entropy production directly implies in our case. 

In addition, the conformal cooling is an energy conserved process, which is fundamentally different from the adiabatic expansion in thermal physics,  where a thermal gas cools during expansion because of performing work on its environment, and hence looses internal energy. For this reason, we denote this unitary evolution as {\em conformal cooling} to distinguish from cooling in more tradition sense.

\subsection{Entropy Production and Conformal Symmetry Breaking}

When the scale symmetry of the interactions is explicitly broken, the broken SO(2,1) symmetry will lead to a source term in the entropy density. Following Appendix \ref{app:entropy}, one finds:

\begin{align}
\left[G_c^{+}\left[\partial_{{\bf r}}, \partial_t\right]  + G_c^{-}\left[\partial_{{\bf r}}, \partial_t\right] \right. &- \left. t^2 \partial_t\right] S({\bf r},t) \nonumber \\
&= \frac{t}{\xi \ln(t)} \frac{\partial}{\partial \xi^{-1}} S({\bf r},t).
\end{align}

\noindent Equivalently, the total entropy will satisfy:

\begin{equation}
\partial_t S(t) =  \frac{1}{\xi t \ln(t)} \frac{\partial}{\partial \xi^{-1}} S(t).
\label{eq:entropy_production}
\end{equation}

\noindent Eq.~(\ref{eq:entropy_production}) implies that $S(t,\xi^{-1}) = S(\ln(t)/\xi)$. For small $\xi^{-1}$, one can expand Eq.~(\ref{eq:entropy_production}) as:

\begin{equation}
\partial_t S(t) = \frac{1}{t \xi} \left. \frac{\partial S}{\partial x}\right|_{x=0} + \frac{\ln(t)}{t \xi^2} \left. \frac{\partial^2 S}{\partial x^2}\right|_{x=0} + ...
\end{equation}

\noindent where $x = \ln(t)/\xi$. For initial conditions which are time reversal invariant, the matrix $\Gamma(0)$ will be an orthogonal matrix. In this case, one can show that the linear term must vanish. As a result the entropy production is proportional to:

\begin{equation}
\partial_t S(t) \approx \frac{\ln(t)}{t \xi^2} \left. \frac{\partial^2 S}{\partial x^2} \right|_{x=0}
\label{eq:ent_rate}
\end{equation}

\noindent As the entropy production will vanish when the Hamiltonian has $SO(2,1)$ conformal symmetry, and the entropy density field is conformal invariant, the entropy production acts as a parameter categorizing conformal symmetry breaking. As one can see, for scale invariant systems, the entropy production rate will vanish, while it must be an explicit function of $t^{\alpha-1}/\xi^{\alpha}$ for systems with explicitly broken scale symmetry.

The long time result should be contrasted to the short time behaviour. For initial conditions that respect time reversal symmetry, the entropy must be an even function of time: $S(t) = S(-t)$. As a result, the short time entropy production rate must be proportional to $t^2$. 

Finally, we note that in the case of isotropic expansion or compressional hydrodynamic flow the entropy production rate can be directly linked to the bulk viscosity, $\zeta_B$, \cite{Landau}.
As a by-product, one can show that the entropy production rate is given by:

\begin{equation}
\partial_t S(t) = -\frac{9}{T_0} \int d{\bf r} \ \zeta_B({\bf r},t).
\end{equation}

\noindent Using this relationship, and Eq.~(\ref{eq:ent_rate}), we obtain a result for the spatially averaged bulk viscosity:

\begin{equation}
\int d{\bf r} \ \zeta_B({\bf r},t) \approx -\frac{T_0}{9} \frac{\ln(t)}{t \xi^2} \left.\frac{\partial^2 S}{\partial x^2} \right|_{x=0}.
\end{equation}

\section{Conclusion}
\label{sec:conclusions}
 
In this work we showed that for Galilean and scale invariant Hamiltonians, the long time dynamics are governed by an emergent conformal symmetry, see Eq.~(\ref{eq:diff_rho_s_3}). This emergent conformal symmetry is robust as it depends on a hidden SO(2,1) symmetry and the resultant invariance of the equation of motion, and is independent of the initial state, which will usually break all the symmetries.

This analysis was carried out using the $m$-body density matrix. The density matrix technique is microscopic, and clearly elucidates the role of the emergent conformal symmetry on the dynamics of atomic systems with scale invariant Hamiltonians. This is in contrast to hydrodynamical phenomenologies which rely on inputs of hydrodynamic coefficients from other separate calculations, or assumptions of the existence of equation of states in non-equilibrium dynamics. Using this approach we examined how the conformal symmetry restricted the dynamics of atomic systems. In particular we examined the dynamics of two physical quantities: the moment of inertia tensor, and the thermodynamic entropy. For the moment of inertia, we showed that the emergent conformal symmetry results in Eq.~(\ref{eq:conformal_moment}). Similarly, the conformal symmetry dictates that the thermal entropy must saturate in the long time limit, Eq.~(\ref{eq:entropy:continuity}).

In addition to these results, we were able to discuss the effect of broken scale invariance near resonance for the three dimensional unitary Fermi gas. In this case, we showed that the dynamics are modified by a non-trivial time dependence of the form $\ln(t)/\xi$. In addition, we showed in Eq.~(\ref{eq:entropy_production}) that there is an one-to-one correspondence between entropy production and broken scale invariance in the long time limit.  

F.Z. is a fellow of Canadian Institute of Advanced Research, and J.M. is in part supported by a fellowship from National Science and Engineering Research Council of Canada (Contract No. 288179).
F.Z. would like to thank Ian Affleck, Leon Balents, Gordon Semenoff, Gora Shlyapnikov, and Congjun Wu for useful discussions.

\appendix

\numberwithin{equation}{section}
\renewcommand\theequation{\Alph{section}.\arabic{equation}}

\section{Invariance of the Equation of Motion}
\label{app:eom}

In this appendix we show that the equation of motion:

\begin{equation}
\partial_t \psi_s({\bf r},t) = i \left[H_s, \psi_s({\bf r},t)\right],
\end{equation}

\noindent is left invariant under scale and conformal transformations. Here $\psi_s({\bf r},t) = e^{i H_s t} \psi({\bf r}) e^{-i H_s t}$ is the time dependent field operator, and $H_s$ is the scale invariant Hamiltonian. The scale invariant Hamiltonian, and the generators of scale (D) and conformal (C) transformations are given by:

\begin{align}
H_s &= \int d{\bf r} \psi^{\dagger}({\bf r}) \left( \frac{p^2}{2} \right) \psi({\bf r}) \nonumber \\
&+ \frac{1}{2} \int d{\bf r} d {\bf r'} \psi^{\dagger}({\bf r})\psi^{\dagger}({\bf r'}) V_s({\bf r - r'}) \psi({\bf r'})\psi({\bf r}) \nonumber \\
 D &= \int d {\bf r} \ \psi^{\dagger}({\bf r}) \left(  {\bf r} \cdot {\bf p} -i \frac{d}{2}\right) \psi({\bf r}) \nonumber \\
C &=  \frac{1}{2} \int d{\bf r} \ r^2 \psi^{\dagger}({\bf r})\psi({\bf r}).
\label{A:operators_def}
\end{align}

\noindent In Eq.~(\ref{A:operators_def}), $V_S({\bf r})$, is a scale invariant two-body potential, and $\psi({\bf r})$ is the second quantized field operators satisfying the anti-commutation relation:

\begin{equation}
\lbrace\psi({\bf r}), \psi^{\dagger}({\bf r'})\rbrace = \delta({\bf r- r'}). 
\end{equation}

\noindent Here we have suppressed the spin indices as they are irrelevant to the following discussions.

Consider a general scale (conformal) transformation. A scale (conformal) transformation is enacted by the unitary operator: $U_{D,(C)}(\lambda) = e^{-i \lambda D (C)}$, where  $\lambda$ is a quantity that parametrizes the extent of the transformation. Using $U_{D, (C)}(\lambda)$, one can show that the transformed equation of motion is given by:

\begin{align}
& \partial_t U_{D,(C)}(\lambda)  \psi_s({\bf r},t) U_{D,(C)}^{\dagger}(\lambda) = \nonumber \\
& i\left[ U_{D,(C)}(\lambda) H_s U_{D,(C)}^{\dagger}(\lambda),   U_{D,(C)}(\lambda) \psi_s({\bf r},t) U_{D,(C)}^{\dagger}(\lambda) \right].
\label{B:eq:transformed_eom}
\end{align}

\noindent  In order to evaluate Eq.~(\ref{B:eq:transformed_eom}), it is necessary to see how both the field operator, as well as the Hamiltonian transforms under scale and conformal transformations. To facilitate this calculation, it is important to note the following commutators:

\begin{align}
[ H_s, C] &= -i D & [D, H_s] &= 2i H_s & [D, C] &= -2 i C,
\label{B:eq:SO21_algebra}
\end{align} 

\noindent Using this algebra, one can show that the field operators transform as:

\begin{align}
U_{D}(\lambda) \psi_s({\bf r},t) U^{\dagger}_{D}(\lambda) &=e^{-\lambda \frac{d}{2}} \psi_s({\bf r}'={\bf r}e^{-\lambda}, t'=t e^{-2\lambda}), 
\nonumber \\ 
U_{C} (\lambda)  \psi_s({\bf r}, t) U^{\dagger}_{C}(\lambda) &=\frac{1}{(1-\lambda t)^{d/2}}\exp(-i\frac{r^2}{2}\frac{\lambda}{1-\lambda t})
\nonumber  \\
& \psi_s({\bf r}'=\frac{\bf r}{1-\lambda t}, t'=\frac{t}{1-\lambda t}),
\label{B:eq:time_dependent_transformation}
\end{align}

\noindent while the Hamiltonian changes as:

\begin{align}
U_D(\lambda) H_s U_D^{\dagger}(\lambda) &= e^{-2 \lambda} H_s \nonumber \\
U_C(\lambda) H_s U_C^{\dagger}(\lambda) &= \frac{1}{(1- \lambda t)^2} H_s + \frac{\lambda}{1-\lambda t}D(t) + \lambda^2 C(t)
\label{B:eq:transformed_H}
\end{align}

\noindent where $H_s$, $C(t)$, and $D(t)$, in the conformally transformed Hamiltonian are written in terms of the operators: $\psi_s({\bf r}',t')$, and ${\bf r}' = {\bf r}/(1-\lambda t)$, $t' = t/(1-\lambda t)$.

Substituting Eqs.~(\ref{B:eq:time_dependent_transformation}) and (\ref{B:eq:transformed_H}) into Eq.~(\ref{B:eq:transformed_eom}) one can show that the equation of motion reduces to:

\begin{equation}
\partial_{t'} \psi_s({\bf r}',t') = i \left[ H_s, \psi_s({\bf r}',t')\right].
\label{B:eq:final_eom}
\end{equation}

\noindent Eq.~(\ref{B:eq:final_eom}) is equivalent to the original, untransformed, equation of motion. Therefore the equations of motion for a quantum system governed by a scale invariant Hamiltonian are invariant to both scale and conformal transformations.

\section{Existence of Conformal Towers}
\label{app:conformal_towers}

In this appendix, we consolidate the discussions of conformal symmetry which were presented in Ref.~\cite{Nishida07}. In particular, we review how the conformal symmetry guarantees that the eigenstates of a quantum system are organized into sets of evenly spaced states.

The conformal symmetry is summarized by the commutators defined in Eq.~(\ref{B:eq:SO21_algebra}). As can be seen the three operators, $H_s$, $C$, and $D$, form a closed group, the conformal, or SO(2,1), group, if the Hamiltonian is rotationally invariant and scale invariant.

Next consider a class of operators that satisfy: $[D,O] = i \Delta_O O$, and $[C,O] = 0$. For our context, this class of operators are called  primary operators. For a full definition and discussion, see Ref.~\cite{Nishida07}. Here we show that:

\begin{equation}
|\psi \rangle = e^{-H_s/\omega}O |vac \rangle,
\end{equation}

\noindent is an eigenstate of the Hamiltonian $H_s + \omega^2 C$, where $|vac \rangle$ is the vacuum state, and $\omega$ is the harmonic trapping frequency. To see that $| \psi \rangle$ is an eigenstate of $H_s + \omega^2 C$, consider:

\begin{align}
(H_s + C)| \psi \rangle =e^{-H_s} (H_s + e^{H_s} C e^{-H_s}) O | vac \rangle,
\end{align}

\noindent where we have set the trap frequency, $\omega$, to unity. Using the identity:

\begin{equation}
e^{H_s} C e^{-H_s} = C + [H_s,C] + \frac{1}{2}[H_s,[H_s,C]] +....
\label{A:eq:identity}
\end{equation}

\noindent the commutation relations contained in Eq.~(\ref{B:eq:SO21_algebra}), and the definition of $O$, one obtains:

\begin{align}
(H_s + C)| \psi \rangle &=e^{-H_s} (C- i D) O | vac \rangle \nonumber \\
&= \Delta_O | \psi \rangle.
\end{align}

\noindent Therefore, $|\psi\rangle$ is an eigenstate of $H_s + C$, with eigenvalue $\Delta_O$. This is true for any primary operator, $O$.

To generate the remaining spectrum of a conformal tower,  we can define the operators:

\begin{equation}
L_{\pm} = H_s - C \pm i D,
\end{equation}

\noindent which satisfy:

\begin{align}
[L_-, L_+ ] &= 4 (H_s+C) & [L_{\pm}, H_s+C] &= 2 L_{\pm}.
\end{align}

These operators $L_{\pm}$ behave identically to the ladder operators for a non-interacting harmonic oscillator, except they raise and lower the energy by two harmonic units.  As can be shown using the above commutation relations, the state $|\psi\rangle$ is the lowest state within a conformal tower. To obtain the remaining states within a conformal tower, one simply needs to apply the raising operator to the state $|\psi \rangle$. 

Another important question is how to label the conformal towers. For this task, it is necessary to find operators that commute with $L_{\pm}$. Two such operators are the total particle number, $N$, and the angular momentum, $L_i$. For this reason, the conformal towers can be labelled by the total particle number of a given conformal tower state, and its angular momentum.

This shows that the spectrum of a conformally symmetric quantum system can be decomposed into a series of conformal towers. The states in each tower are evenly spaced by two harmonic units.

\section{Conformal Towers and Dynamics}
\label{app:conformal_tower_dynamics}

In this appendix we show the use of applying the conformal tower states to the study of dynamics. Consider a given state: $|\psi_0 \rangle$ that is unitarily time evolved by a nearly scale invariant Hamiltonian, $H = H_s + V$, where $H_s$ is the scale invariant Hamiltonian, and $V$ is some explicit perturbation that breaks the scale symmetry. The dynamics of the quantum state is given by:

\begin{align}
|\psi(t) \rangle &= e^{-i H t} |\psi_0 \rangle \nonumber \\
&= e^{-i H_s t} e^{i H_s t}e^{-i H t} |\psi_0 \rangle \nonumber \\
&= e^{-i H_s t} U(t) |\psi_0 \rangle,
\end{align}

\noindent where we have defined:

\begin{align}
U(t) &= e^{i H_S t} e^{-i H t} \nonumber \\
&= T \exp\left(\int_0^t dt' \ V_I(t') \right).
\label{B:eq:U}
\end{align}

\noindent In Eq.~(\ref{B:eq:U}) $T$ is the time ordering operator and $V_I(t)$ is the deviation from scale invariance in the interaction picture:

\begin{equation}
V_I(t) = e^{i H_s t} V e^{-i H_s t}.
\label{B:eq:V}
\end{equation}

Now insert a complete set of eigenstates for the Hamiltonian $H_s+ C$, $|O^l_n \rangle$, the many body conformal tower states:

\begin{equation}
|\psi(t) \rangle = \sum_{n,l} e^{-i H_s t} |O_n^l \rangle \langle O_n^l | U(t) | \psi_0 \rangle.
\label{B:eq:general_evolution}
\end{equation}

The value of this basis can be seen by examining the state $|O_n^l(t) \rangle  = e^{-i H_s t} |O_n^l \rangle$. Following the previous discussions, we act the Hamiltonian, $H_s + C$, on this state to show:

\begin{align}
e^{-i H_s t} (H_s + C)|O_n^l \rangle &=E_{n^l} |O_n^l (t) \rangle \nonumber \\
&=e^{-i H_s t} (H_s + C) e^{i H_s t} |O_n^l(t) \rangle
\end{align}

\noindent This can once again be solved exactly by using the commutation relationships in Eq.~(\ref{B:eq:SO21_algebra}):

\begin{align}
e^{-i H_s t} (H_s + C)|O_n^l \rangle &= \left((1+t^2)H_s -t D + C\right) |O_n^l(t) \rangle,
\end{align}

\noindent Although this may look like an arbitrary operator we note that under the substitution:

\begin{align}
\tilde{{\bf p}} &= \sqrt{1+t^2}({\bf p} - \frac{{\bf r} t}{1+t^2}) & \tilde{{\bf r}} &= \frac{{\bf r}}{\sqrt{1+t^2}},
\label{B:eq:coordinates}
\end{align}

\noindent one obtains the final result:

\begin{align}
e^{-i H_s t} (H_s + C) | O_n^l \rangle  &= (\tilde{H}_s + \tilde{C}) |O_n^l(t) \rangle = E_n^l |O_n^l(t) \rangle.
\label{B:eq:result}
\end{align}

Eq.~(\ref{B:eq:result}) states that the eigenstates of $H_s + C$, when time evolved by the scale invariant Hamiltonian, $H_s$, will be instantaneous eigenstates of a harmonic oscillator that is defined in terms of the coordinates in Eq.~(\ref{B:eq:coordinates}).

The usefulness of the conformal tower states is that their dynamics are perfectly adiabatic, i.e. a trivial time dependent rescaling. For example, if you prepare a scale invariant system in a conformal tower state, the problem can be seen as static in terms of the coordinates defined in Eq.~(\ref{B:eq:coordinates}). This is exactly the situation for a non-interacting Gaussian wave packet. The wave function maintains its shape, although it is a large superposition of plane wave states.

For a more physical interpretation of these conformal tower states, it was shown in Ref.~\cite{Maki18} that the dynamics could be understood in an expanding, non-inertial, reference frame, or comoving frame for short.  In this frame, a fictitious harmonic oscillator potential is present. The eigenstates of the Hamiltonian in the comoving frame are nothing more than the conformal tower states.

For systems that break scale invariance, it is necessary to also consider the matrix elements of $V_I(t)$ on conformal tower states:

\begin{align}
\langle m | V_I(t) |n \rangle = \langle m(t) | V | n(t) \rangle.
\end{align}

\noindent This will produce a time dependence to the interaction.  It was shown in Ref.~\cite{Maki18}, that the effects of this explicit breaking of scale invariance on the long time dynamics can be evaluated for three dimensional Fermi gases near resonance. In this case one finds:

\begin{equation}
U(t \gg 1) = e^{-i \frac{1}{\xi} \ln(t) \tilde{V}}
\end{equation}

\noindent where $\xi$ is the scattering length, and $\tilde{V}$ is a dimensionless, universal, matrix, that depends only on the number of particles.

\section{Structure of the Density Matrix}
\label{app:density_matrix}

In this appendix, we derive the properties of the density matrix for a quantum system governed by a Hamiltonian that is either or nearly scale invariant. Here we consider the $m$-body density matrix:

\begin{align}
P_m(\lbrace {\bf r}_{1i} \rbrace, & \lbrace {\bf r}_{2i} \rbrace,t ) = \nonumber \\
& \langle \psi_0 | \prod_{i=1}^m \psi^{\dagger}({\bf r}_{2i},t) \prod_{i=1}^m\psi({\bf r}_{1i},t) | \psi_0 \rangle,
\label{D:eq:density_matrix}
\end{align}

\noindent where $i =1,2,...N$, and the field operators are defined as:

\begin{equation}
\psi({\bf r},t) = e^{i H t} \psi({\bf r}) e^{-i H t}.
\end{equation}

\noindent Here we note that the Hamiltonian is given by:

\begin{equation}
H = H_s + V,
\end{equation}

\noindent where $H_s$ is the resonant scale invariant Hamiltonian, and $V$ is an explicit symmetry breaking term.

Again it is possible to use the interaction picture to separate the scale invariant dynamics from the dynamics governed by the symmetry breaking terms:

\begin{align}
P_m&( \lbrace {\bf r}_{1i} \rbrace, \lbrace {\bf r}_{2i} \rbrace,t ) = \nonumber \\
& \langle \psi_0 |U^{\dagger}(t) \prod_{i=1}^m\psi_s^{\dagger}({\bf r}_{2i},t) \prod_{i=1}^m \psi_s({\bf r}_{1i},t) U(t)| \psi_0 \rangle,
\end{align}

\noindent where $U(t)$ is defined in Eq.~(\ref{B:eq:U}), and the field operators are now defined in terms of the scale invariant Hamiltonian:

\begin{equation}
\psi_s({\bf r},t) = e^{i H_s t} \psi({\bf r}) e^{-i H_s t}.
\label{A2:eq:si_psi}
\end{equation}

Inserting two complete sets of conformal tower states allows one to separate the scale invariant motion from the terms that break the scale invariance:

\begin{align}
P_m(\lbrace {\bf r}_{1i} \rbrace, \lbrace {\bf r}_{2i} \rbrace, t) &= \sum_{n,l;n',l'} \rho_{n,n'}^{l,l'}(\lbrace {\bf r}_{1i} \rbrace, \lbrace {\bf r}_{2i} \rbrace, t) \Gamma_{n',n}^{l',l}(t) \nonumber \\
\Gamma_{n',n}^{l',l}(t) &= \langle O_{n'}^{l'} | U(t) | \psi_0 \rangle \langle \psi_0 | U^{\dagger}(t) | O_n^l\rangle \nonumber \\
\rho_{n,n'}^{l,l'}(\lbrace{\bf r}_{1i} \rbrace, \lbrace {\bf r}_{2i} \rbrace, t) &= \nonumber \\
&\langle O_{n}^l | \prod_{i=1}^m \psi_s^{\dagger}({\bf r}_{2i},t) \prod_{i=1}^m \psi_s({\bf r}_{1i},t) | O_{n'}^{l'} \rangle.
\label{A2:eq:def}
\end{align}

Let us begin by focusing on the scale invariant piece, $\rho_{n,n'}^{l,l'}(\lbrace {\bf r}_{1i} \rbrace, \lbrace {\bf r}_{2i} \rbrace, t)$. We can use the fact that the conformal tower states, $| O_n^l\rangle$, are eigenstates of $H_s + C$, where the trap frequency is set to unity, to obtain the following identity:

\begin{align}
&\rho_{n,n'}^{l,l'}(\lbrace {\bf r}_{1i} \rbrace, \lbrace {\bf r}_{2i} \rbrace, t) = e^{-i(E_n^l - E_{n'}^{l'}) \lambda} \cdot \nonumber \\
&\langle O_n^l | e^{i(H_s + C)\lambda} \prod_{i=1}^m\psi_s^{\dagger}({\bf r}_{2i},t) \prod_{i=1}^m\psi_s({\bf r}_{1i},t) e^{-i(H_s +C) \lambda} |O_{n'}^{l'} \rangle.
\end{align}

\noindent This transformation should not effect the form of $\rho_{n,n'}^{l,l'}(\lbrace {\bf r}_{1i} \rbrace, \lbrace {\bf r}_{2i} \rbrace, t)$ as it is nothing more than the identity. Therefore, we can obtain the condition:

\begin{align}
&\frac{\partial}{\partial \lambda } \rho_{n,n'}^{l,l'}(\lbrace {\bf r}_{1i} \rbrace, \lbrace {\bf r}_{2i} \rbrace, t) = 0 \nonumber \\
&= -i(E_n-E_m) \rho_{n,n'}^{l,l'}(\lbrace {\bf r}_{1i} \rbrace, \lbrace {\bf r}_{2i} \rbrace, t) \nonumber \\
&+ i \langle O_n^l|  e^{i H_s t}\left[ H_s + C(t),\prod_{i=1}^m \psi^{\dagger}({\bf r}_{2i}) \prod_{i=1}^m \psi({\bf r}_{1i})  \right] e^{-i H_s t} | O_{n'}^{l'} \rangle, \nonumber \\
\label{A2:eq:rho_s_eta}
\end{align}

\noindent where 

\begin{align}
C(t) &= e^{-i H_s t} C e^{i H_s t} \nonumber \\
&= C - D t + t^2 H_s.
\end{align}

\noindent Using Eq.~(\ref{B:eq:SO21_algebra}), and the commutators, 

\begin{align}
[D, \psi({\bf r})] &= i \left({\bf r} \cdot \partial {\bf r} + \frac{d}{2} \right) \psi({\bf r}) \nonumber \\
[C, \psi({\bf r})] &= - \frac{r^2}{2} \psi({\bf r}) \nonumber \\
[D, \psi^{\dagger}({\bf r})] &= i \left({\bf r} \cdot \partial {\bf r} + \frac{d}{2} \right) \psi^{\dagger}({\bf r}) \nonumber \\
[C, \psi^{\dagger}({\bf r})] &= \frac{r^2}{2} \psi^{\dagger}({\bf r}),
\end{align}

\noindent it is possible to evaluate Eq.~(\ref{A2:eq:rho_s_eta}), and to obtain a differential equation that $\rho_{n,n'}^{l,l'}(\lbrace {\bf r}_{1i} \rbrace, \lbrace {\bf r}_{2i} \rbrace, t)$ must satisfy:

\begin{align}
0 &= \left[(1+t^2)\partial_t + t\sum_{i=1}^m \left( {\bf r}_{1i} \cdot \partial {\bf r}_{1i} + {\bf r}_{2i} \cdot \partial {\bf r}_{2i}+d \right)  \right. \nonumber \\
&+ \left. i \sum_{i=1}^m \frac{ r_{2i}^2-r_{1i}^2}{2} -i(E_n^l-E_{n'}^{l'})\right] \rho_{n,n'}^{l,l'}(\lbrace {\bf r}_{1i} \rbrace, \lbrace {\bf r}_{2i} \rbrace, t).
\label{A2:eq:diff_rho_s}
\end{align}

In the long time limit, the last term of Eq.~(\ref{A2:eq:diff_rho_s}) can be neglected. The remaining terms are nothing more than the generator of conformal transformations. As a result, $\rho_{n,n'}^{l,l'}$ is an eigenfunction of the conformal generator with zero eigenvalue. Therefore the scale invariant piece of the density matrix will be invariant under conformal transformations:

\begin{align}
\rho_{n,n'}^{l,l'}&(\lbrace {\bf r}_{1i} \rbrace, \lbrace {\bf r}_{2i} \rbrace,t) = e^{i C \lambda} \rho_{n,n'}^{l,l'}(\lbrace {\bf r}_{1i} \rbrace, \lbrace {\bf r}_{2i} \rbrace,t) e^{- i C \lambda} \nonumber \\
&=\frac{1}{(1-\lambda t)^{dm}}  e^{\frac{i}{2} \sum_i (r_{2i}^2 - r_{1i}^2) \frac{\lambda}{1- \lambda t}} e^{-i (E_n^l-E_{n'}^{l'}) \arctan(t)} \nonumber \\
& \rho_{n,n'}^{l,l'}\left(\lbrace \frac{{\bf r}_{1i}}{1- \lambda t} \rbrace, \lbrace \frac{{\bf r}_{2i}}{1 - \lambda t} \rbrace,\frac{t}{1- \lambda t}\right),
\end{align}

\noindent Consequently, the full $m$-body density matrix for a scale invariant system will satisfy:

\begin{align}
0 &= \left[ t^2\partial_t + t \sum_{i=1}^m \left( {\bf r}_{1i} \cdot \partial {\bf r}_{1i} + {\bf r}_{2i} \cdot \partial {\bf r}_{2i} +d \right)  \right. \nonumber \\
&+ \left. i \sum_{i=1}^m \frac{r_{2i}^2-r_{1i}^2}{2}\right] P_m(\lbrace {\bf r}_{1i} \rbrace, \lbrace {\bf r}_{2i} \rbrace, t) \nonumber \\
&- i \sum_{n,l; n',l'} (E_n^l-E_{n'}^{l'}) \rho_{n,n'}^{l,l'}(\lbrace {\bf r}_{1i} \rbrace, \lbrace {\bf r}_{2i} \rbrace, t) \Gamma_{n',n}^{l',l}(0).
\label{A3:eq:diff_eqn_rho_res}
\end{align}

\noindent To obtain the above result, note that $\Gamma_{n',n}^{l',l}
(t)$ is time independent for scale symmetric Hamiltonians, see Eq.~(\ref{A2:eq:def}). Taking the long time limit, one can show that Eq.~(\ref{A3:eq:diff_eqn_rho_res})  reduces to the generator of conformal symmetry:

\begin{align}
0&=\left[ t^2 \partial t+ t \sum_{i=1}^m \left( {\bf r}_{1i} \cdot \partial_{{\bf r}_{1i}} + {\bf r}_{2i} \cdot \partial_{{\bf r}_{2i}} + d \right) \right. \nonumber \\
& \left. + i \sum_{i=1}^m \frac{r_{2i}^2 - r_{1i}^2 }{2} \right] P_N(\lbrace {\bf r}_{1i} \rbrace,\lbrace {\bf r}_{2i} \rbrace,t)
\label{A3:eq:generator_conformal}
\end{align}

\noindent As a result, the $m$-body density matrix will be an eigenfunction of the generator of conformal transformations with zero eigenvalue, and will be a conformally invariant function.

With the dynamics of the scale invariant piece of the density matrix determined, we can now address the dynamics of a nearly scale invariant system. Such a study is tantamount to examining the matrix, $\Gamma_{n',n}^{l',l}(t)$, defined in Eq.~(\ref{A2:eq:def}).  This matrix must satisfy the following differential equation:

\begin{equation}
\partial_t \Gamma(t) = -i \left[ V_I(t), \Gamma(t) \right],
\end{equation}

\noindent where $V_I(t)$ was defined in Eq.~(\ref{B:eq:V}). 

As discussed in Ref.~\cite{Maki18}, and the previous appendix, for nearly resonant, three dimensional Fermi gases, the matrix $U(t)$, defined in Eq.~(\ref{B:eq:U}), has the long time form:

\begin{equation}
U(t) = e^{-i \frac{1}{\xi} \ln(t) \tilde{V}},
\end{equation}

\noindent where, again, $\tilde{V}$ is a universal, dimensionless, time independent matrix that depends only on the number of particles, and $\xi$ is the scattering length. 

This explicit form of $U(t)$ means that the differential equation for $\Gamma(t)$ will have the form:

\begin{equation}
\frac{\partial}{\partial (\ln(t)/\xi)}\Gamma(t) = -i\left[ \tilde{V}, \Gamma(t) \right].
\label{A2:eq:gamma_eom}
\end{equation}

\noindent Eq.~(\ref{A2:eq:gamma_eom}) implies that the Gamma matrix is a function of $\ln(t)/\xi$:

\begin{equation}
\Gamma(t) = \Gamma \left( \frac{\ln(t)}{\xi} \right).
\end{equation}

\noindent As a result, one can show:

\begin{equation}
t^2 \partial_t \Gamma(t) = \frac{t}{\xi \ln(t)} \frac{\partial}{\partial \xi^{-1}} \Gamma(t).
\label{A2:eq:gamma_relation}
\end{equation}

\noindent Eq.~(\ref{A2:eq:gamma_relation}) allows one to write a differential equation for the total $N$-body density matrix near resonance:

\begin{align}
0 &= \left[ t^2 \partial_t + t \sum_{i=1}^m \left( {\bf r}_{1i} \cdot \partial {\bf r}_{1i} + {\bf r}_{2i} \cdot \partial {\bf r}_{2i} +d \right)  \right. \nonumber \\
&+ \left. i \sum_{i=1}^m \frac{r_{2i}^2-r_{1i}^2}{2} - \frac{t}{\xi \ln(t)} \frac{\partial}{\partial \xi^{-1}}\right] P_m(\lbrace {\bf r}_{1i} \rbrace, \lbrace {\bf r}_{2i} \rbrace, t) \nonumber \\
&- i \sum_{n,l; n',l'} (E_n^l-E_{n'}^{l'}) \rho_{n,n'}^{l,l'}(\lbrace {\bf r}_{1i} \rbrace, \lbrace {\bf r}_{2i} \rbrace, t) \Gamma_{n',n}^{l',l}(t).
\label{A3:eq:diff_eqn_rho_nearly}
\end{align}

\noindent The difference between Eqs.~(\ref{A3:eq:diff_eqn_rho_nearly}) and (\ref{A3:eq:diff_eqn_rho_res}) is the presence of a source term which depends on the derivative of the $m$-body density matrix with respect to the inverse of the correlation length: $\xi^{-1}$.

\section{Compressional and Elliptic Flow}
\label{app:flow}

In this appendix we give explicit expressions for  the leading long time correction for the moment of inertia tensor in the case of compressional and elliptic flow of a strongly interacting Fermi gas in three spatial dimensions. The moment of inertia tensor is defined as:

\begin{equation}
I_{i,j}(t) = \int d {\bf r} \ r_i r_j P_1({\bf r},{\bf r},t),
\end{equation}

\noindent where $P_1({\bf r},{\bf r},t)$ is the one body density matrix defined in Eq.~(\ref{D:eq:density_matrix}). As shown in the preceding appendix, for scale invariant systems, there is a hidden SO(2,1) symmetry that restricts the dynamics of the density matrix, see Eq.~(\ref{A3:eq:diff_eqn_rho_res}). Using Eq.~(\ref{A3:eq:diff_eqn_rho_res}) we can obtain a differential equation for the moment of inertia tensor:

\begin{align}
0&=\left[(1+t^2) \partial_t -  2t\right] I_{i,j}(t)\nonumber \\
&- t^2\sum_{n,l;n',l'} i\left(E_n^l -E_{n'}^{l'}\right) \int d{\bf r} \ r_i r_j \ \rho_{n,n'}^{l,l'}({\bf r},{\bf r},0) \Gamma_{n',n}^{l',l}(0), \nonumber \\
\label{A4:eq:differential_eqn_moment}
\end{align}

\noindent where we have neglected the symmetry breaking terms, and have used the emergent conformal symmetry to write:

\begin{equation}
\rho_{n,n'}^{l,l'}({\bf r},{\bf r},t) \approx \frac{1}{t^d}e^{-i(E_n^l-E_{n'}^{l'}) \arctan(t)}\rho_{n,n'}^{l,l'}\left( \frac{{\bf r}}{t},\frac{{\bf r}}{t},0 \right).
\end{equation}




\subsection{Isotropic expansion Flow}

In compressional flow, the dynamics are isotropic, therefore we focus on the dynamics of $\langle r^2 \rangle(t) = \sum_i I_{i,i}(t)$. The conformal symmetry requires:

\begin{equation}
\langle r^2 \rangle(t) = \left( v^2 t^2 + A t + B\right) \langle r^2 \rangle(0).
\end{equation}

\noindent In this case one finds:

\begin{itemize}
\item $A = 0$
\item $B = v^2 - \frac{1}{2}F_0$ 
\end{itemize}

\noindent where:

\begin{align}
F_0 = \frac{1}{\langle r^2 \rangle(0)} & \sum_{n,n',l} (-1)^{n-n'} 4 (n-n')^2  \nonumber \\
& \int d{\bf r} \ r^2 \rho_{n,n'}^{l,l}({\bf r}, {\bf r}, 0) \Gamma_{n',n}^{l',l}(0).
\end{align}

\subsection{Elliptic Flow}

In elliptic flow, there are both isotropic and anisotropic contributions to the dynamics. Here we focus on the anisotropic contributions, which are encapsulated in the quadrupole moment:

\begin{equation}
\langle Q_{i,j} \rangle (t) = \int d{\bf r} \ (3r_i r_j -  r^2 \delta_{i,j}) P_1({\bf r}, {\bf r},t).
\end{equation}

\noindent The conformal symmetry requires:

\begin{equation}
\langle Q_{i,j} \rangle(t) = \left( v^2_{i,j} t^2 + A_{i,j} t + B_{i,j}\right).
\end{equation}

In this section, we focus on the contributions from the  the total moment of inertia during the expansion dynamics. In this case one finds:

\begin{itemize}
\item $A_{i,j} = F_2^{i,j}$
\item $B_{i,j} = v_{i,j}^2 - \frac{1}{2}F_0^{i,j}$
\end{itemize}

\noindent where:

\begin{eqnarray}
F_2^{i,j} &=& \sum_{n,l'; n',l'} (E_n^l-E_{n'}^{l'}) \sin\left((E_n^l - E_{n'}^{l'})\frac{\pi}{2} \right) \nonumber \\
& \cdot & \int d{\bf r} \ (3 r_i r_j- r^2 \delta_{i,j})  \  \rho_{n,n'}^{l,l'}({\bf r},{\bf r},0) \Gamma_{n',n}^{l',l}(0) \nonumber \\
F_0^{i,j} &=& \sum_{n,l'; n',l'} (E_n^l-E_{n'}^{l'})^2 \cos\left((E_n^l - E_{n'}^{l'})\frac{\pi}{2} \right) \nonumber \\
& \cdot &\int d{\bf r} \ (3 r_i r_j- r^2 \delta_{i,j}) \  \rho_{n,n'}^{l,l'}({\bf r},{\bf r},0) \Gamma_{n',n}^{l',l}(0). \nonumber \\
\end{eqnarray}

\section{The Thermodynamic Entropy}
\label{app:entropy}

In this appendix we examine the thermal entropy for scale invariant systems. For a Fermi gas consisting of $N$ particles, the thermodynamic entropy can be written as:

\begin{align}
S(t)  &=-Tr\left[ P_N \log P_N \right] \nonumber \\
&= -\prod_{i=1}^N\int d{\bf r}_{1,i} \ d{\bf r}_{2,i} P_N(\lbrace {\bf r}_{1,i} \rbrace, \lbrace {\bf r}_{2,i} \rbrace,t) \nonumber \\
& \cdot \log P_N (\lbrace {\bf r}_{2,i} \rbrace, \lbrace {\bf r}_{1,i} \rbrace,t) ,
\end{align}

\noindent where $P_N$ is the $N$-body density matrix, defined in Eq.~(\ref{D:eq:density_matrix}), the integrals are over the $2N$ coordinates, and $\log P_N$ is the logarithm of the density matrix, defined as:

\begin{align}
P_N(\lbrace {\bf r}_{1i} \rbrace, \lbrace {\bf r}_{2i} \rbrace,t) & \equiv  e^{\log P_N} (\lbrace {\bf r}_{1i} \rbrace, \lbrace {\bf r}_{2i} \rbrace,t)  \nonumber \\
&=\sum_n \frac{1}{n!} (\log P_n) ^n (\lbrace {\bf r}_{1i} \rbrace, \lbrace {\bf r}_{2i} \rbrace,t)
\label{A3:eq:def_log}
\end{align}

\subsection{Entropy Conservation for Scale Invariant Interactions}

We first consider the case where the system is evolving under a scale invariant Hamiltonian, $H_s$.  In this case, the long time asymptotic behaviour of the density matrix is governed by the emergent conformal invariance, see Eq.~(\ref{A3:eq:generator_conformal}). However, in order to understand the role of conformal symmetry on the entropy, it is necessary to consider how the logarithm of the $N$-body density matrix transforms under a conformal transformation. Under a conformal transformation, Eq.~(\ref{A3:eq:def_log}) transforms as:

\begin{align}
\tilde{P}_N=U_C(\lambda)& P_N(\lbrace {\bf r}_{1i} \rbrace, \lbrace {\bf r}_{2i} \rbrace,t) U_C^{\dagger}(\lambda) = \nonumber \\
& \sum_n \frac{1}{n!} U_C(\lambda)X^n (\lbrace {\bf r}_{1i} \rbrace, \lbrace {\bf r}_{2i} \rbrace,t) U_C^{\dagger}(\lambda)
\label{A3:eq:transformed_log_def}
\end{align}

\noindent where we use the abbreviation: $X = \log P_N$, for this section.

Let us focus on the right hand side of this equation. Expanding out the product of matrices, one can show that:

\begin{equation}
U_C(\lambda) X^n U_C^{\dagger}(\lambda) = \left(U_C(\lambda) X U_C^{\dagger}(\lambda)\right)^n
\label{A3:eq:log_rho_n}
\end{equation}


\noindent Using the definition of the conformal transformation, the transformed logarithm of the density matrix will be given by:

\begin{align}
\tilde{X}=U_C(\lambda)X (\lbrace {\bf r}_{1i} \rbrace &, \lbrace {\bf r}_{2i} \rbrace,t) U_C^{\dagger}(\lambda) = \nonumber \\
&\frac{1}{(1-\lambda t)^{dN}} e^{\frac{i}{2} \sum_{i=1}^N ({\bf r}_{2i}^2 - {\bf r}_{1i}^2) \frac{\lambda}{1- \lambda t}}  \nonumber \\
&X\left( \frac{\lbrace {\bf r}_{1i} \rbrace}{1 - \lambda t}, \frac{\lbrace {\bf r}_{2i} \rbrace}{1 - \lambda t}, \frac{t}{1- \lambda t} \right). \nonumber \\
\label{A3:eq:log_rho_conformal}
\end{align}

\noindent It is also straightforward to show that the $n$th power of $X$ will transform in the exact same way as $X$ itself.

Now that we know how the $n$th power of $X$ transforms, it is possible to examine Eq.~(\ref{A3:eq:transformed_log_def}). Under the conformal transformation, we note that:

\begin{align}
\tilde{P}_N &=e^{\tilde{X}},    & \tilde{X} &=\log \tilde{P}_N.
\label{eq:f6}
\end{align}

\noindent Now we note that the $N$-body density matrix, $P_N$, is a conformally invariant function, i.e. $\tilde{P}_N=P_N$. Hence, following Eq.~(\ref{eq:f6}), $\tilde{X} = X$. Since $X$ is also a conformally invariant function, it must satisfy the same differential equation as $P_N$, see Eq.~(\ref{A3:eq:generator_conformal}):

\begin{align}
0 &= \left[ t^2 \partial_t + t \sum_{i=1}^N \left( {\bf r}_{1i} \cdot \partial {\bf r}_{1i} + {\bf r}_{2i} \cdot \partial {\bf r}_{2i} +d \right)  \right. \nonumber \\
&+ \left. i \sum_{i=1}^N \frac{{\bf r}_{2i}^2-{\bf r}_{1i}^2}{2}\right] X(\lbrace {\bf r}_{1i} \rbrace, \lbrace {\bf r}_{2i} \rbrace, t).
\label{A3:eq:diff_eqn_log_rho}
\end{align}

Since both the density matrix, and it's logarithm are eigenfunctions of the generator of conformal transformations with zero eigenvalue, it is possible to obtain a differential equation for the entropy. Consider the thermal entropy density defined as:

\begin{align}
S({\bf r},t) &= -\prod_{i =1}^N \int d {\bf r}_{1i} \int d{\bf r}_{2i} \delta({\bf r}_{1i=1} - {\bf r})  \nonumber \\
&P_N(\lbrace {\bf r}_{1i} \rbrace, \lbrace {\bf r}_{2i} \rbrace,t) \log P_N(\lbrace {\bf r}_{2i} \rbrace, \lbrace {\bf r}_{1i} \rbrace,t).
\end{align}

\noindent Using Eqs.~(\ref{A3:eq:diff_eqn_rho_res}) and (\ref{A3:eq:diff_eqn_log_rho}), it is possible to derive an equation of motion for the entropy density:

\begin{align}
0= [t^2 \partial_t + t ({\bf r} \cdot \partial {\bf r} + d) ]S({\bf r},t).
\label{A3:eq:diff_eqn_S}
\end{align}

\noindent This can be re-written in a more convenient form as:

\begin{equation}
\partial_t S({\bf r},t) + \partial {\bf r} \cdot \left( \frac{{\bf r}}{t} S({\bf r},t) \right) =0.
\label{A3:eq:continuity}
\end{equation}

\noindent Eq.~(\ref{A3:eq:continuity}) is nothing more than the conservation equation for the entropy density. As a result, the total thermal entropy must be conserved:

\begin{equation}
0 = \partial_t S(t), 
\end{equation}

\noindent i.e. conformal invariance means entropy is conserved in the long time limit.

\subsection{Entropy Production for Systems with Interactions Breaking Scale Symmetry}

As we have seen in the previous section, there is no entropy production in the long time limit for scale invariant interactions. In this section we consider the case when the scale symmetry is broken explicitly. As we have discussed in Appendix \ref{app:density_matrix}, the differential equation for the density matrix, and equivalently, the logarithm to the density matrix, will be given by Eq.~(\ref{A3:eq:diff_eqn_rho_nearly}).  The main difference between the differential equation at resonance, Eq.~(\ref{A3:eq:diff_eqn_rho_res}) and Eq.~(\ref{A3:eq:diff_eqn_rho_nearly}) is the presence of a source term.

The presence of this source term will naturally lead to the production of entropy. For three dimensional nearly resonant Fermi gases, the equation for the entropy density becomes:

\begin{equation}
\left[ t^2 \partial_t + t ({\bf r} \cdot \partial {\bf r} + d)- \frac{t}{\xi \ln(t)} \frac{\partial}{\partial \xi^{-1}} \right] S({\bf r},t).
\end{equation}

\noindent Equivalently, the total entropy will satisfy:

\begin{equation}
\partial_t S(t) =  \frac{1}{\xi t \ln(t)} \frac{\partial}{\partial \xi^{-1}} S(t).
\end{equation}

\end{document}